\RequirePackage{fix-cm}
\documentclass[smallcondensed,final]{svjour3}
\usepackage{epsfig,graphicx,mathptmx,amssymb}

\begin{document}

\title{\centering{DEBRIS DISK SIZE DISTRIBUTIONS: STEADY STATE COLLISIONAL EVOLUTION WITH
P-R DRAG AND OTHER LOSS PROCESSES}}
\titlerunning{Debris disk size distributions}

\author{\centering{M. C. WYATT$^1$, C. J. CLARKE$^1$ and M. BOOTH$^{1,2,3}$ \\
    \it{$^1$ Institute of Astronomy, University of Cambridge, Madingley Road, Cambridge CB3 0HA, UK
    \\\email{wyatt@ast.cam.ac.uk} \\ 
    $^2$ Herzberg Institute of Astrophysics, 5071 West Saanich Road, Victoria, B. C., V9E 2E7, Canada \\
    $^3$ University of Victoria, Finnerty Road, Victoria, B. C., V8W 3P6, Canada}}}
\institute{}

\authorrunning{M.C.Wyatt et al.}

\date{Submitted: 26 January 2011, Revised: 1 March 2011}

\maketitle

\begin{abstract}
We present a new scheme for determining the shape of the size distribution, and its evolution,
for collisional cascades of planetesimals undergoing destructive collisions and loss
processes like Poynting-Robertson drag.
The scheme treats the steady state portion of the cascade by equating mass loss and gain in
each size bin;
the smallest particles are expected to reach steady state on their collision timescale,
while larger particles retain their primordial distribution.
For collision-dominated disks, steady state means that mass loss rates in logarithmic
size bins are independent of size.
This prescription reproduces the expected two phase size distribution, with ripples above
the blow-out size, and above the transition to gravity-dominated planetesimal strength.
The scheme also reproduces the expected evolution of disk mass, and of dust mass,
but is computationally much faster than evolving distributions forward in time.
For low-mass disks, P-R drag causes a turnover at small sizes to a size distribution
that is set by the redistribution function (the mass distribution of fragments produced in
collisions).
Thus information about the redistribution function may be recovered by measuring
the size distribution of particles undergoing loss by P-R drag, such as that traced by
particles accreted onto Earth.
Although cross-sectional area drops with age $\propto t^{-2}$ in the PR-dominated regime,
dust mass falls $\propto t^{-2.8}$, underlining the importance of understanding which
particle sizes contribute to an observation when considering how disk detectability evolves.
Other loss processes are readily incorporated;
we also discuss generalised power law loss rates, dynamical depletion, realistic radiation
forces and stellar wind drag.
\end{abstract}

\section{Introduction}
\label{s:intro}
Many nearby main sequence stars are seen to be surrounded by $\mu$m-sized dust known as
debris disks (see review in Wyatt 2008).
The dust is short-lived indicating that larger planetesimals are present that feed the
observed dust (Backman \& Paresce 1993).
It is thought that the large planetesimals are destroyed in mutual collisions, creating fragments
that also collide to produce yet smaller fragments, and so on until the dust is small enough
to be removed from the system by radiation pressure;
i.e., that there is a continuous size distribution extending from small to large objects
known as a collisional cascade.

In this respect extrasolar debris disks are similar to the asteroid and Kuiper belts in
the Solar System.
The size distribution of large objects in the Solar System's belts is well characterised
observationally (Gladman et al. 2009; Fuentes et al. 2010).
There remains debate as to the origin of this distribution at the largest sizes
(Durda et al. 1998; Bottke et al. 2005; Morbidelli et al. 2009), but it is recognised
that these populations have undergone, and continue to undergo, collisional evolution
which means that their size distributions should also extend down to dust sizes.
Exactly how the distributions extrapolate down is less well constrained
(e.g., Schlichting et al. 2009).
However, dust is seen in the inner Solar System that migrates inwards past the Earth due to
Poynting-Robertson drag (Leinert \& Gr\"{u}n 1990; Dermott et al. 2001), though it
is debated as to whether this dust has an asteroidal or cometary origin
(Durda \& Dermott 1997; Nesvorn\'{y} et al. 2010).

One big difference with extrasolar debris disks is that, to be detected above the light from
the star, their dust has to be orders of magnitude more abundant than dust in the Solar System.
This means that it has been possible to ignore the effects of {P-R} drag when studing
their evolution, since for such disks collisional timescales (that scale inversely with total
dust mass) are much shorter than P-R drag timescales (that are independent of dust mass)
(Wyatt 2005).
The regime of such \textit{collision-dominated} disks is now (relatively) well understood.
Early analytical models treated systems in which the size distribution is defined by a single power
law, and showed that disk luminosity is expected to stay constant, but to fall off inversely with
age at late times (Dominik \& Decin 2003; Wyatt et al. 2007a), in a manner consistent
with the observed debris disk population (Wyatt et al. 2007b).
More recent analytical models allow for three phases in the size
distribution to allow for planetesimals having a size-dependent dispersal threshold (L\"{o}hne
et al. 2008; Heng \& Tremaine 2009), and provide a better fit to the size distribution
expected from more detailed numerical models that solve for the evolution of an
input size distribution given assumptions about the outcome of collisions between different
sized particles (Krivov et al. 2006; Th\'{e}bault \& Augereau 2007).
However, the analytical models cannot fit details in the expected size distribution,
which is not just defined by power laws, and neither do they consider effects other than
collisions.

There is now a need to understand how loss processes affect the size distribution, and so the
detectability, of a debris disk.
Current instrumentation is probing dust levels at which drag may start to become
important, and as searches for Earth-analogs intensify, it is inevitable that lower levels
of dust will be probed and the dynamics in this regime needs to be properly understood
(Beichman et al. 2006).
It is also becoming clear that stellar wind drag, which acts in a similar manner to
P-R drag, may be important for young stars with high mass loss rates (Plavchan et al. 2005;
Reidemeister et al. 2010).

Although the competition between collisions and P-R drag has been considered in the Solar
System (Leinert et al. 1983; Gr\"{u}n et al. 1985; Ishimoto 2000), such studies focus on explaining
the currently observed dust distribution, and are not readily applicable to earlier epochs or
to the debris disks of other stars.
Collisional models that do study the long-term evolution of the asteroid belt typically exclude small
particles and focus on fitting the large body size distribution (Bottke et al. 2005),
whereas models for structures in the zodiacal cloud often only treat the dynamics of individual
particles essentially ignoring collisions (Dermott et al. 1994; Grogan et al. 2001).
However, more recently the balance of collisions and P-R drag has been incorporated into codes
that evolve size distributions with time and so far results have been presented for two systems
(Vitense et al. 2010; Reidemeister et al. 2010).
Other models are also being developed that incorporate a (less detailed) prescription for
collisions into N-body simulations (Stark \& Kuchner 2009; Kuchner \& Stark 2010).

Here we present a new scheme that can be used to determine the size distribution of 
a planetesimal belt undergoing destructive collisions as well as loss processes.
This scheme is not expected to provide a more faithful measure of the size distribution
than would be found by evolving a distribution forward in time.
Rather its value is in the speed with which the distribution can be determined
(in seconds), with its ability to reproduce details in the steady state size distribution,
including when loss processes are acting, and with the simplicity of the analytical
presentation which permits an analytical understanding of how various processes affect
the size distribution.
Although the scheme presented here is in its most basic form, considering the
1-dimensional size distribution (rather than, e.g., including spatial dimensions), 
and treating collisional destruction with a 1-dimensional redistribution function
(essentially averaging the outcome over all possible collisions), extension to
extra dimensions is possible in principle.
\S \ref{s:col} outlines the development of the scheme by application to the collision-dominated
regime, then \S \ref{s:pr} continues by detailing the effect of P-R drag;
\S \ref{s:gen} considers other loss processes, and conclusions are given in \S \ref{s:conc}.

\section{Steady state in the collisional regime}
\label{s:col}
Consider a belt of planetesimals in which $m_{k}$ is the total
mass in the $k$-th bin, where $D_{k}$ is the typical size of
planetesimals in the bin, and bins are logarithmically spaced in
size, with $k=1$ being the largest bin and working to smaller
sizes with increasing $k$ so that $D_{k+1}/D_{k}=1-\delta$ where
$\delta$ is approximately the logarithmic bin width, since it is
also assumed that $\delta \ll 1$.
Collisions between planetesimals cause the mass to be redistributed
amongst the bins, and furthermore loss processes may cause mass
to be removed from the belt, and there may be sources that add mass
to the belt.

From mass conservation for the $k$-th bin
\begin{equation}
  \dot{m}_{k} = \dot{m}_{k}^{+c} - \dot{m}_{k}^{-c} - \dot{m}_{k}^{-l} + \dot{m}_{k}^{+g},
  \label{eq:mdotk}
\end{equation}
where the superscripts $+c, -c, -l, +g$ refer to the mass gained from collisions
in other bins, the mass lost from collisions in this bin, the mass removed
by loss processes, and the mass gained from other sources, respectively.
Note that our 1-dimensional prescription of the belt does not have a
radial dimension and so considers loss processes such as P-R drag as a sink,
rather than as a diffusive process (e.g. Gorkavyi et al. 1997).
For now we consider the situation where $\dot{m}_{k}^{+g}=\dot{m}_{k}^{-l}=0$, but
return to loss processes in \S \ref{s:pr}.

\subsection{Mass loss is independent of size}
\label{ss:mk}
We define the redistribution function $f(i,k)$ to be the fraction of mass leaving
the $i$-th bin from collisions that goes into the $k$-th bin.
Since mass must be conserved $\sum_{k=1}^{\infty} f(i,k) = 1$.
In this paper the model is being applied to situations in which collisions
are destructive, so that $f(i,k \leq i)=0$ and $\sum_{k=i+1}^{\infty} f(i,k) = 1$,
though it may be possible to extend this to situations where accretion
can also take place (see, e.g., Birnstiel et al. 2011).

Let us further assume that the redistribution function is scale independent, so
that $f(i+n,k+n)=f(i,k)$, where $n$ is any integer, which we can implement by defining
a new redistribution function needing just one parameter $F(k-i)=f(i,k)$.
Putting this into equation~(\ref{eq:mdotk}) and giving collisional gains and
losses in terms of the redistribution function we find
\begin{equation}
  \dot{m}_{k} = \sum_{i=1}^{k-1} \dot{m}_{i}^{-c} F(k-i) - \dot{m}_{k}^{-c}.
\end{equation}

If we assume that the size distribution is in quasi-steady state, so that
$\dot{m}_{k} \approx 0$, and append $s$ to any subscripts of quantities that are
calculated in steady state, we find that
\begin{equation}
  \dot{m}_{ks}^{-c} = \sum_{i=1}^{k-1} \dot{m}_{is}^{-c} F(k-i).
  \label{eq:ss1}
\end{equation}
Since $\sum_{l=1}^{\infty} F(l) = 1$, then as long as
$\sum_{l=k}^{\infty} F(l) \ll 1$,
one solution to equation~(\ref{eq:ss1}) must be that
\begin{equation}
  \dot{m}_{ks}^{-c} = \dot{m}_{is}^{-c} = C,
  \label{eq:ss2}
\end{equation}
where $C$ is a constant;
i.e., the mass loss rate in logarithmic bins is independent of size.

Note that the main assumption to arrive at this result was
that the redistribution function is scale independent.
However, there is a further assumption regarding whether the
size distribution is infinite in extent (e.g., Dohnanyi 1969).
Truncated distributions are considered in more detail in following
sections, but here we note that for distributions truncated at large
sizes, equation~(\ref{eq:ss2}) is expected to apply at sizes for which the mass
that would have been replenished from sizes above the truncation is
negligible (i.e., it applies for $k \gg 1$ and redistribution functions
that are weighted toward larger sizes).

\subsection{Mass loss rate}
\label{ss:ststsizedist}
Equations~(\ref{eq:ss1}) and (\ref{eq:ss2}) can be used to determine the
steady state size distribution, $m_{ks}$.
To do so, we consider that the mass loss rate from the $k$-th bin due to
collisions is
\begin{equation}
  \dot{m}_{k}^{-c} = m_{k} R_{k}^{c},
  \label{eq:mdotk-c}
\end{equation}
where $R_{k}^{c}$ is the rate at which individual objects of size $D_{k}$
undergo catastrophic collisions.

A catastrophic collision is defined as one in which the mass of the largest
remnant after the collision has less than half of the mass of the original
object.
This definition is used for our \textit{typical} collision, not because we
believe that catastrophic collisions are the only important collisions ---
indeed it has been shown that cratering collisions with much lower energy may
also result in significant mass loss (e.g., Kobayashi \& Tanaka 2010) ---
but because we expect the overall mass loss rate (i.e., including both
catastrophic and cratering collisions) for all bins to scale with the
rate of catastrophic collisions within them, at least for distributions
that are infinite in extent. 
It may be that mass loss in fact scales with the rate of collisions in which
the largest remnant has slightly more or less than 50\% the mass of the original
object.

The catastrophic collision rate for a planetesimal of size $D_{k}$
can be determined using the particle-in-a-box approach and integrating 
over the size distribution of impactors that have sufficient specific
incident energy ($Q$) to cause a catastrophic collision, which is defined as
the dispersal threshold $Q_{D}^{\star}$.
For collisions that occur at relative velocity $v_{rel}$ the smallest
impactors that cause catastrophic destruction are those of size $X_{c}D_{k}$,
where
\begin{equation}
  X_{c}=(2Q_{D}^{\star}/v_{rel}^2)^{1/3},
  \label{eq:xc}
\end{equation}
which we denote in the following equation as those with an index $i_{ck}$.
Thus the collision rate, both in its discrete and continuous forms, is
\begin{eqnarray}
  R_{k}^{c} & = & \sum_{i=1}^{i_{ck}} \frac{3m_{i}}{2\rho \pi D_{i}^3}
                  (D_{k}+D_{i})^2 P_{ik}, \label{eq:rkc1} \\
            & = & \int_{X_{c}D_{k}}^{D_1} n(D_{i})
                  0.25 (D_{k}+D_{i})^2 P_{ik} dD_{i}, \label{eq:rkc2}
\end{eqnarray}
where $\rho$ is particle density, $P_{ik}$ is the intrinsic collision probability
used by other authors (e.g., Wetherill 1967) between particles $i$ and $k$ that
is equal to $\pi v_{rel}/V$ for all particle sizes for applications in this paper,
and $V$ is the volume through which the planetesimals are moving, $n(D)$ is the
number of planetesimals in the size distribution per unit diameter, and the
gravitational focussing factor has been neglected for this calculation.

\subsection{Analytical steady state method}
\label{sss:analyt}
The continuous form of the collision rate (equation~\ref{eq:rkc2}) can be used
to determine the steady state size distribution analytically, for certain
assumptions.
By integrating the mass distribution $(\pi \rho D^3/6) n(D)$ over the bin
we find that for small $\delta$
\begin{equation}
  m_{k} \approx (\pi\rho/6) n(D_{k}) D_{k}^4 \delta,
  \label{eq:mk}
\end{equation}
and so
\begin{equation}
  \dot{m}_{k}^{-c} \propto n(D_{k}) D_{k}^4 \int_{X_{c}D_{k}}^{D_1}
                      n(D_{i})(D_{k}+D_{i})^2 dD_{i}.
  \label{eq:mdotk_ndxcdk}
\end{equation}
Further assuming that
\begin{equation}
  n(D) = K D^{-\alpha},
  \label{eq:nd}
\end{equation}
and that $\alpha > 3$ and $X_{c} \ll 1$ so that collisions close to the catastrophic collision
threshold dominate the rate (i.e., so we can take only the term with $D_{k}^2$
in the integral and ignore the upper limit), gives
\begin{equation}
  \dot{m}_{k}^{-c} \propto X_{c}^{1-\alpha} D_{k}^{7-2\alpha}.
  \label{eq:mdotk_xcdk}
\end{equation}
If $Q_{D}^{\star} \propto D^b$ then we find that
\begin{equation}
  \dot{m}_{k}^{-c} \propto D_{k}^{7-2\alpha+(1-\alpha)b/3},
  \label{eq:mdotk_dk}
\end{equation}
and for this to be independent of size as required for steady state (equation~\ref{eq:ss2})
\begin{equation}
  \alpha = (7+b/3)/(2+b/3).
  \label{eq:alpha}
\end{equation}
This reduces to the canonical $\alpha=3.5$ distribution for the situation where
strength is independent of size (Dohnanyi 1969; Tanaka et al. 1996), and replicates
the distribution found analytically by other authors when strength depends on size
(O'Brien \& Greenberg 2003; L\"{o}hne 2008; L\"{o}hne et al. 2008; Heng \& Tremaine 2009;
Birnstiel et al. 2011; Belyaev \& Rafikov 2011).

\subsection{Numerical steady state method}
\label{sss:num}
The discrete form of the collision rate (equation~\ref{eq:rkc1}) can also be used to
determine the steady state size distribution numerically.

\subsubsection{Method 1}
\label{sss:method1}
If we impose a constant mass loss rate per logarithmic bin, as argued to be the case
in \S \ref{ss:mk}, then we can rearrange equation~(\ref{eq:mdotk-c}) and combine with
equation~(\ref{eq:ss2}) to get
\begin{equation}
  m_{ks} = C/R_{k}^{c}.
  \label{eq:mks_num1}
\end{equation}
This can be solved iteratively, and converges quickly to find the steady state
distribution.
The shape of the resulting distribution is independent of the total mass in
the distribution, because multiplying the whole distribution
$m_{ks}$ by $N$ would result in the mass loss in the $k$-th bin
$\dot{m}_{k}^{-c}$ and the mass gain $\dot{m}_{k}^{+c}$ both increasing by $N^2$.
Application of this method is discussed in \S \ref{sss:num1appl}.

\subsubsection{Method 2}
\label{sss:method2}
Alternatively we can relax the assumption of equation~(\ref{eq:ss2}) and apply
a similar method more directly to equation~(\ref{eq:ss1}), and so equate
mass loss with mass gain in each bin by iterating
\begin{eqnarray}
  m_{ks} & = & C_{k} / R_{k}^{c},   \label{eq:mks_num2} \\
  C_{k}  & = & \sum_{i=1}^{k-1} \dot{m}_{is}^{-c} F(k-i). \label{eq:ck}
\end{eqnarray}
In this case, by the arguments of \S \ref{ss:mk}, we would expect to end up with
a solution for which $C_k$ is approximately constant (at least for most particle
sizes).

To implement this method requires the redistribution function $F(l)$ to be defined
(see \S \ref{ss:redistrib}).
Without further constraints this method would converge to a distribution with zero mass.
This is because the lack of larger particles means that the mass input rate to the
top bin is zero, which can only be balanced by having zero mass there;
if that is the case then the mass input to the bin below this is zero, and so on.
This issue can be remedied by constraining the largest bin(s) to
have either a fixed mass distribution, or a fixed mass input rate (e.g.,
by adding $\dot{m}_{0}^{+g}$ to $C_k$).
Further numerical tricks are also required to achieve convergence.
If numeric subscripts indicate the iteration number, or more specifically
a variable calculated using the mass distribution from that iteration,
then we found convergence with the following two-step scheme
\begin{eqnarray}
  m_{ks1.5} & = & [(C_{k1}/R_{k1}^c)   + m_{ks1}]/2, \\
  m_{ks2}   & = & [(C_{k1}/R_{k1.5}^c) + m_{ks1}]/2.
\end{eqnarray}
This method is employed in \S \ref{ss:evoln}, and further extended to
incorporate loss due to P-R drag in \S \ref{s:pr}.

\subsection{Redistribution function}
\label{ss:redistrib}
Where it is necessary in this paper, we assume that the fragments 
from the destruction of objects of size $D_{k}$ follow the power law of
equation~(\ref{eq:nd}), with parameters identified by subscript $r$,
from sizes of $\eta_{rmax}D_{k}$ down to infinitesimally small particles.
Unless otherwise stated we consider $\eta_{rmax}=2^{-1/3}$ (i.e.,
the largest objects have half the mass of the original), which corresponds
to catastrophic collisions dominating the redistribution of mass.
In this paper we also consider a function with $\eta_{rmax}=1$, which
corresponds to cratering collisions dominating mass redistribution.
For $\eta_{rmax}<1$ we also include a small but finite fraction of mass
in particles in the $\eta_{rmax}D_{k}$ to $D_{k}$ size range using a steep power
law (with $\alpha=-20$)\footnote{This is necessary to allow the size
distribution to be calculated at late times in \S \ref{ss:evoln} when
$D_t>\eta_{rmax}D_{1}$.}.
The slope is always assumed to have $\alpha_{r}<4$ so that the mass in fragments
is weighted toward large objects.
These assumptions mean that the fractional mass distribution of these
fragments is
$\bar{m}(D) \approx (4-\alpha_r)D^{3-\alpha_r}(\eta_{rmax}D_k)^{\alpha_r-4}$
across the appropriate range.
Integrating over the size range of bin $D_{k+l}$, and assuming $\delta \ll 1$,
gives
\begin{equation}
  F(l) \approx \eta_{rmax}^{\alpha_r-4}(4-\alpha_r)\delta(1-\delta)^{l(4-\alpha_r)}.
\end{equation}

Although it is the redistribution function $F(k-i)$ that is employed throughout this
paper, it is worth considering a more general form of the redistribution function,
since the techniques presented herein may also applicable in this instance.
Most numerical models of collisional evolution define the redistribution function as
$f_{2}(i,j,k)$, the fraction of mass leaving the $i$-th bin from collisions with the
$j$-th bin that goes into the $k$-th bin.
In general such redistribution functions also have some form of scale independence, and
can, e.g., be written as $F_{2}(j-j_{ci},k-i)$, where $j_{ci}$ is the index of the size
that causes catastrophic destruction of objects of size $D_i$.
This is because a collision with the $j_{ci}$-th bin ends up, by definition, with a
largest remnant that has half the mass of the original, and collisions with other bins
end up with largest remnants that scale as a function of the ratio $Q/Q_D^\star$, and
so have redistribution functions that only depend on $j-j_{ci}$.

If we assume that $F_{2}(j-j_{ci},k-i)$ is the true redistribution function, then we 
have implicitly averaged over $j$, weighted by the frequency of such collisions, to
get $F(k-i)$ as the redistribution of mass in a \textit{typical} collision.
Thus we can write
\begin{equation}
  F(k-i) = \frac{ \sum_{j=1}^{\infty} m_j (D_i+D_j)^2 D_j^{-3} F_{2}(j-j_{ci},k-i) } 
                { \sum_{j=1}^{j_{ci}} m_j (D_i+D_j)^2 D_j^{-3} }.
  \label{eq:fkifull}
\end{equation}
This illustrates that the $F(k-i)$ redistribution function may not be scale
independent, even if $F_2(j-j_{ci},k-i)$ is, because it includes information about the
size distribution (see also Belyaev \& Rafikov 2011).
In particular, it may be expected to be different at locations where there are sharp
transitions in the size distribution.
These issues may also contribute to the timescale of mass loss in a \textit{typical}
collision being different to the timescale for catastrophic collisions, as discussed in
\S \ref{ss:ststsizedist}.
Nevertheless, we retain the $F(k-i)$ redistribution function for this paper,
since it is expected to be valid over most of the size distribution, and is shown in
later sections to be in agreement with more detailed models.

\subsection{Application to realistic dispersal laws}
\label{ss:strengthgravity}
As reported by several authors studying collision outcomes (e.g., Durda et al. 1998;
Benz \& Asphaug 1999), the size dependence of the dispersal threshold can be expressed as 
\begin{equation}
  Q_{D}^{\star} = Q_{a} D^{-a} + Q_{b} D^{b},
  \label{eq:qd}
\end{equation}
where the subscripts $a$ and $b$ refer to contributions from the planetesimal's
material strength and from its self-gravity, respectively, and $a$ and $b$ are
positive.
The weakest planetesimals then have size
\begin{equation}
  D_{w} = \left( \frac{aQ_{a}}{bQ_{b}} \right)^{1/(a+b)};
  \label{eq:dw}
\end{equation}
e.g., for planetesimals made of basalt impacting at 3 km s$^{-1}$ (Benz \& Asphaug 1999),
reasonable values are $Q_{a}=790$ J kg$^{-1}$, $a=0.38$, $Q_{b}=0.017$ J kg$^{-1}$,
$b=1.36$, for which $D_{w}=0.23$ km.
Here we adopt slightly different values of $Q_a=620$ J kg$^{-1}$, $a=0.3$,
$Q_b=5.6 \times 10^{-3}$ J kg$^{-1}$, $b=1.5$, to allow comparison with the
results of L\"{o}hne et al. (2008) in later sections.

\subsubsection{Analytical 2-phase distribution}
\label{sss:2phase}
A reasonable approximation is that the only contribution to $Q_{D}^\star$
is from the strength component for $D \ll D_{w}$ (the strength regime) and from the gravity 
component for $D \gg D_{w}$ (the gravity regime).
Application of the analysis of \S \ref{sss:analyt} then suggests that the
size distribution should have two phases described by power laws (equation~\ref{eq:nd})
with 
\begin{equation}
  \alpha_a = (7-a/3)/(2-a/3),
  \label{eq:alphaa}
\end{equation}
in the strength regime and $\alpha_b=\alpha$ given by equation~(\ref{eq:alpha}) in the
gravity regime.
Power law indices in this paper, like $\alpha_a$, that refer to equation~(\ref{eq:nd})
are summarised in Table~\ref{tab:tab2}, along with a summary of diameters that define
where these apply.
For the dispersal law assumed here, $\alpha_a=3.63$ and $\alpha_b=3.0$.

\begin{table}
  \begin{center}
    \caption{Summary of (some of) the parameters used in the paper:
    power law indices that refer to equation~(\ref{eq:nd}), and diameters.
    Diameters (e.g., $D_x$) also have corresponding indices (e.g., $i_x$)
    to denote bin number.}
    \begin{tabular}{ccc}
      \hline
      \hline
      Power law index     & Regime                         & Equation \\
      \hline
      $\alpha_p$          & Primordial                     & given \\
      $\alpha_b$          & Collisional (gravity)          & \ref{eq:alpha} \\
      $\alpha_a$          & Collisional (strength)         & \ref{eq:alphaa} \\
      $\alpha_{pr}$       & PR-dominated                   & \ref{eq:alphapr} \\
      $\alpha_l$          & General loss-dominated         & \ref{eq:alphal} \\
      $\alpha_r$          & Redistribution function        & given \\
      \hline
      \hline
      Diameter            & Meaning (or transition)        & Equation \\
      \hline
      $D_1$               & Largest planetesimals          & given \\
      $D_t$               & Primordial-collisional         & \ref{eq:dt} \\
      $D_w$               & Strength-gravity scaling       & \ref{eq:dw} \\ 
      $D_{pr}$            & Collisional-PR-dominated       & \ref{eq:dprdef} \\
      $D_{l}$             & Collisional-loss-dominated     & \ref{eq:dldef} \\
      $D_{bl}$            & Radiation pressure-dominated   & \ref{eq:dbl} \\
      \hline
      \hline
    \end{tabular}
    \label{tab:tab2}
  \end{center}
\end{table}

Although one might expect the distribution to be continuous at $D_{w}$, closer
inspection shows that this cannot be the case, since the condition given by
equation~(\ref{eq:ss2}) should hold regardless of the dispersal law, and so
should hold on both sides of (and at) the transition.
The discontinuity at $D_{w}$ was pointed out by O'Brien \& Greenberg (2003).
The relative heights of the two distributions, and so the magnitude of the
discontinuity, can be determined by including the constants
of proportionality in equation~(\ref{eq:mdotk_dk}).
These constants are
$\propto K_a^2 (1-\alpha_a)^{-1} (v_{rel}^2/2)^{\alpha_a/3}Q_a^{(1-\alpha_a)/3}$
for the strength regime, and likewise except with subscript $_b$ for the gravity
regime, resulting in
\begin{equation}
  \frac{K_a}{K_b} = \sqrt{\frac{1-\alpha_a}{1-\alpha_b}}
                    \left( \frac{v_{rel}^2}{2} \right)^{(\alpha_b-\alpha_a)/6}
                    \left( \frac{Q_b^{(1-\alpha_b)/6}}{Q_a^{(1-\alpha_a)/6}} \right).
  \label{eq:kakb}
\end{equation}

\begin{figure}
  \vspace{-0.1in}
  \begin{center}
    \begin{tabular}{c}
      \hspace{-0.0in} \includegraphics[height=2.3in]{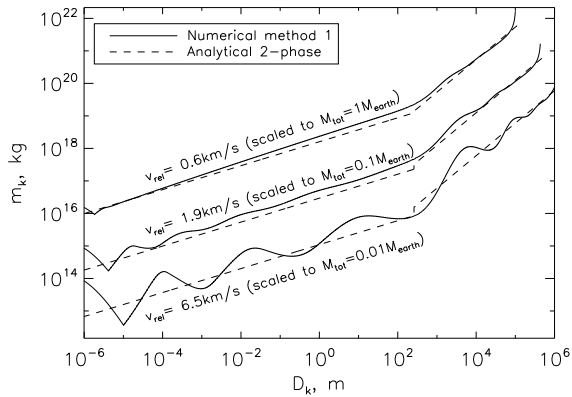} \\[-0.1in]
    \end{tabular}
    \caption{Steady state size distribution for particles between $1\mu$m and
    1000km undergoing catastrophic collisions.
    The y-axis is the mass in logarithmic bins centred on sizes given on the x-axis,
    where bin width is given by $\delta=0.01$. 
    Three distributions are shown for relative velocities of 0.6, 1.9, 6.5 km $s^{-1}$
    using the same law for the dispersal threshold.
    The distributions are scaled to total masses of 0.01, 0.1 and 1 times
    the mass of the Earth to allow these to be distinguished (though the shape of
    the distribution is independent of total mass).
    The distributions are calculated using the analytical 2-phase (dashed line)
    and numerical method 1 (solid line).}
    \label{fig:fig1}
  \end{center}
\end{figure}

Scaling the total mass in the 2-phase distribution to that required, say $M_{tot}$,
means that the distribution is now completely defined.
In Figure~\ref{fig:fig1} we show with a dashed line the predicted distributions for
the assumed dispersal law for different values of relative velocity.
Note that we did not change the dispersal law with collision velocity, although this
might be expected (Benz \& Asphaug 1999; Stewart \& Leinhardt 2009).
The total mass in the distributions were scaled to allow them to be distinguished.
The main thing to note from the analytical distribution (for now) is that the
discontinuity is bigger for larger relative velocities, as expected from
equation~(\ref{eq:kakb}).

It should also be remarked that for low collision velocities, $X_{c}$ can be
greater than 1 for the largest particles, so that our simplistic view
of collisions in \S \ref{ss:ststsizedist} implies zero mass loss rate for some
particles near the top end of the distribution.
In practise these particles can still lose mass through cratering collisions,
illustrating how our assumption that mass loss rates scale only with the catastrophic
collision rate breaks down in certain regions.
Such particles could not reach steady state if our assumptions were valid,
and a more detailed treatment is needed to assess evolution in this regime
(e.g., \S \ref{ss:redistrib}).
Rather than assume some primordial distribution for such particles, here
we set the mass to zero in all bins for which $X_{c}>1$ at the top
end.
Note that, although $X_{c}$ can also be greater than 1 at small sizes,
this does not prevent the destruction of such grains, since they can be
destroyed by particles larger than themselves.

\subsubsection{Numerical 2-phase with ripples}
\label{sss:num1appl}
There were many approximations in the analytical 2-phase model that
are overcome in the numerical model.
Notably the numerical model should be valid regardless of the
shape of the dispersal law, and should also be applicable (within the
limitations of the assumptions) in the situation where
the size distribution is finite in extent, being truncated at small
and/or large sizes.

Figure~\ref{fig:fig1} also shows with solid lines the result of
numerical method 1 (\S \ref{sss:method1}) that solves equation~(\ref{eq:mks_num1}).
Convergence of this scheme was confirmed by checking that $\dot{m}_{k}^{-c}$ is
indeed the same for all bins, as expected by construction.
The first thing to note is the excellent agreement with the 2-phase analytical
model.
There is slight quantitative disagreement on the slope of the distributions, which
we attribute to the terms that were discarded from equation~(\ref{eq:mdotk_ndxcdk})
when deriving equation~(\ref{eq:mdotk_xcdk}).
Thus, from Figure~\ref{fig:fig1}, we expect the slope of the mass distribution
in the strength regime to be slightly steeper than predicted by $\alpha_a$, and for
the distribution in the gravity regime to be slightly shallower than that predicted
by $\alpha_b$.

As expected for a realistic distribution, there is no discontinuity at $D_{w}$.
However, as discussed by O'Brien \& Greenberg (2003, see their Fig.~3)
and previously noted by Durda et al. (1998), the transition at $D_{w}$ does cause
a ripple that extends to larger sizes, and is captured by the numerical model.
The transition does not affect the distribution in the strength regime because,
as discussed in O'Brien \& Greenberg (2003), the distribution in the strength
regime is independent of that in the gravity regime, which follows in our
analysis because $\dot{m}_{k}^{-c}$ depends almost entirely on the number of
smaller objects in the distribution with size around $X_{c}D_{k}$.
We find that the magnitude of the ripple (i.e., the departure from the power
law slope) depends on the collision velocity, with higher velocities causing
larger ripples.
The ripple decreases in magnitude to larger sizes, with a peak-to-peak
wavelength that also depends on collision velocity and varies with size
(decreasing wavelength to larger sizes).
All such effects follow naturally from the dependence of $X_{c}$ with size,
which is the parameter that sets the magnitude and wavelength of the ripple,
since for particles of a given size the larger $X_{c}$ is the nearer in size
particles have to be to affect their number.
In the gravity regime $X_{c}$ increases with size (and velocity) causing
the observed dependence.
This also means there should be a dependence on $Q_{D}^\star$ in that weaker
planetesimals should also result in a more pronounced ripple with a longer
wavelength.

A similar ripple is also seen at small sizes, caused by the fact that the
distribution is truncated at $1\mu$m.
While this truncation is implemented simply by not considering bins $<1$ $\mu$m, a
truncation is physically motivated by radiation pressure, since grains smaller
than $D_{bl}$, where
\begin{equation}
  D_{bl} \approx (2300/\rho)(L_\star/M_\star)
  \label{eq:dbl}
\end{equation}
in $\mu$m (for $\rho$ in kg m$^{-3}$, and $L_\star$ and $M_\star$ in Solar units),
are placed on hyperbolic trajectories as soon as they are created (e.g., Burns et al. 1979).
In this case the lack of $<1\mu$m impactors causes the equilibrium number of $1\mu$m grains
to be enhanced.
The consequence of this is an enhanced destruction rate of objects of sizes typically
destroyed by $1\mu$m grains (i.e., those for which $X_cD=1\mu$m).
This causes a dearth of such grains and so a higher equilibrium number of grains that
would have been destroyed by them, and so on.
Such ripples have previously been discussed by Campo-Bagatin et al. (1994) and seen
in the simulations (Th\'{e}bault et al. 2003; Krivov et al. 2006).
There is a similar dependence on collision velocity of the magnitude of the
ripple, and its wavelength, in that both increase with $v_{rel}$.
This can again be understood in terms of $X_{c}$, which is both smaller for larger velocities
(and so particles can be affected by those at much smaller relative sizes), and
which decreases with size in the strength regime.
This ripple would also be expected to be more pronounced for weaker planetesimals.
The details of this ripple should, however, not be overinterpreted.
For example, if the truncation is caused by radiation pressure, then grains just
above the blow-out limit (equation~\ref{eq:dbl}) should also have a higher
eccentricity than larger particles, which numerical simulations show 
would affect the shape of this ripple (Krivov, Mann \& Krivova 2000; Th\'{e}bault \&
Augereau 2007).
The ripple may also be less prominent if particles have a range of compositions
and so a range of truncation sizes and strengths.

In short, all of the features of the steady state size distribution arising from
collisions that have been discussed in the literature (power laws and ripples)
are reproduced with our simple numerical scheme.
This allows the shape of the distribution to be analysed without having to evolve
size distributions forward in time until they reach steady state, which can be
time-consuming for large particles with low collision rates, if timesteps are set
by the small particles that evolve much faster.
This should allow us to consider how the steady state distribution depends on
parameters such as the shape of the dispersal threshold law and the collision
velocity.
But more importantly for this paper, allows us to consider how that shape is
affected by loss processes.

\subsection{Timescale to damp perturbations from steady state}
\label{ss:truncationlarge}
The steady state size distributions presented in the preceding sections 
have some issues at the top end of the distribution.
It was already discussed that the numerical scheme employed, to keep
mass loss rate independent of size, should be valid for most of the size
range, but that for a distribution that is truncated at sizes above
$D_1$ this should be invalid for the size range in which the redistribution
function means that objects of size $D_1$ (and larger, had they existed) deposit
a significant fraction of their mass.
Although a numerical scheme was outlined that should be able to
account for this (equation~\ref{eq:mks_num2}), we should consider the validity
of the assumption that the largest particles are in steady state.

First consider a bin to have a mass that departs from that expected when the
distribution is in quasi-steady state by a small fraction $\epsilon_{k}$, so that
\begin{equation}
  m_{k} = m_{ks} (1+\epsilon_{k}).
  \label{eq:mkpert}
\end{equation}
Since $\dot{m}_{k}^{+c}$ and $R_{k}^{c}$ are independent of $m_{k}$
for small enough bins (or a small enough perturbation), substituting
equation~(\ref{eq:mkpert}) into equation~(\ref{eq:mdotk}) gives
\begin{equation}
  \dot{\epsilon}_{k}/\epsilon_{k} = -R_{k}^{c}.
  \label{eq:epsk}
\end{equation}
In other words, perturbations are damped on the collision timescale.

\subsection{Analytical 3-phase distribution}
\label{sss:an3ph}
Another way of interpreting equation~(\ref{eq:epsk}) is that the collision
timescale also defines the timescale for a given bin size to reach steady
state.
Thus for a given evolution timescale, $t_{age}$, all particles with collision
rates higher than $1/t_{age}$ (i.e., all of the smallest particles) would
be expected to have reached steady state, which we denote with subscript $ss$,
whereas those with smaller collision
rates (i.e., the largest particles) would be expected to retain their
primordial size distribution, which we define here using equation~(\ref{eq:nd})
with a subscript $p$.
We define $D_{t}$ to denote those particles with index $k=i_t$ for which
\begin{equation}
  R_{k}^{c}(D_t) = 1/t_{age}.
  \label{eq:dt}
\end{equation}

This motivated L\"{o}hne et al. (2008) to define a 3-phase size distribution
comprised of the 2 phases in steady state given in \S \ref{sss:2phase},
in addition to the primordial phase.
These authors connected the size distribution in each phase continuously.
However, if it is assumed that there are two distinct portions of the size distribution,
primordial and steady state, with the destruction of primordial particles 
feeding those in steady state (in a manner similar to the
approach of Dominik \& Decin 2003), then there may be a discontinuity at this transition.

The reason is that bins in the steady state population have mass input equal to
that lost.
This means that the rate of mass loss from the steady state population
is the sum over this population considering just the mass put into particles
small enough to fall outside the cascade
\begin{equation}
  \dot{m}^{-}_{ss} = C \sum_{i=i_t}^{i_{min}} \sum_{k=i_{min}}^\infty F(k-i),
  \label{eq:mdot-cc}
\end{equation}
where we have used the fact that $\dot{m}^{-c}_{is}=C$ in the regime considered to take
it outside of the sum, and $i_{min}=i_{bl}$ for initial considerations in this paper, but
would be the larger of $i_{bl}$ and $i_{pr}$ if a consideration of P-R drag was necessary
(\S \ref{s:pr}).
Since $C$ scales with the square of the total mass in steady state (for fixed $i_t$), so
does the mass loss rate.
The mass input to the collisional cascade, however, is set by the
sum over all bins in the primordial distribution, considering just the mass
put into sizes in the collisional cascade
\begin{equation}
  \dot{m}^{+}_{ss} = \sum_{i=1}^{i_t} \dot{m}_{i}^{-c} \sum_{k=i_t}^{i_{min}} F(k-i),
  \label{eq:mdot+cc}
\end{equation}
where $\dot{m}_{i}^{-c}$ is set mostly by the primordial distribution.

Thus in this approximation, the mass in the steady state population
represents a balance between the two, i.e., $\dot{m}_{ss}^{+}=\dot{m}_{ss}^{-}$ so that
\begin{equation}
  C = \frac{\sum_{i=1}^{i_t} \dot{m}_{i}^{-c} \sum_{k=i_t}^{i_{min}} F(k-i)}
           {\sum_{i=i_t}^{i_{min}} \sum_{k=i_{min}}^\infty F(k-i)}
  \label{eq:ccc}
\end{equation}
and does not necessarily lead to a continuous distribution at $D_t$.
This is particularly so if $D_t$ is close to $D_1$, since at this point the mass
input to the steady state population from primordial particles is low,
and so that population must also have low mass and the discontinuity is large.

In fact the transition from primordial to steady state is not expected to be
discontinuous, and such a discontinuity may compromise the validity of the
scale independence assumption of our redistribution function (\S \ref{ss:redistrib}).
Nevertheless it is hoped that this approximation still captures the important physics.
For example, although the discontinuity is large in the regime in which $D_t$ approaches
$D_1$, the primordial population may, in some respects such as the long term evolution
of the steady state population, be irrelevant at this point.
This is because, in \S \ref{ss:evoln}, the system age is set to be equal to the
collisional lifetime of objects of size $D_t$, which may well have a negligible
contribution from the primordial population.
If this is the case, then the collisional lifetime would be expected to scale
inversely with the mass in the steady state population, leading to disk mass falling off
inversely with age, as expected from previous analytical and numerical studies
(e.g., Dominik \& Decin 2003; Wyatt et al. 2007a; L\"{o}hne et al. 2008).

\subsection{Evolving the size distribution}
\label{ss:evoln}
The scheme of \S \ref{sss:an3ph} allows us, for a given primordial distribution
and transition size, to define the discontinuity and so derive a 3 phase 
distribution analytically.
Further equating the collision lifetime of objects of size $D_t$ with the system
age would also permit the evolution of an input primordial size distribution to be
determined, in a similar manner to L\"{o}hne et al. (2008).
Here we follow a comparable approach, except that the 3-phase distribution is 
derived using numerical method 2 (\S \ref{sss:method2}; i.e., using equation \ref{eq:mks_num2}),
with the largest size bins (above $D_t$) fixed at the primordial size distribution.
This automatically captures the discontinuity (and its effect on the collision
rate of particles at the transition) and achieves steady state in the
appropriate size range.

\begin{figure*}
  \begin{center}
    \vspace{-0.0in}
    \begin{tabular}{cccc}
      \hspace{-0.15in} \textbf{(a)} &
      \hspace{-0.5in} \psfig{figure=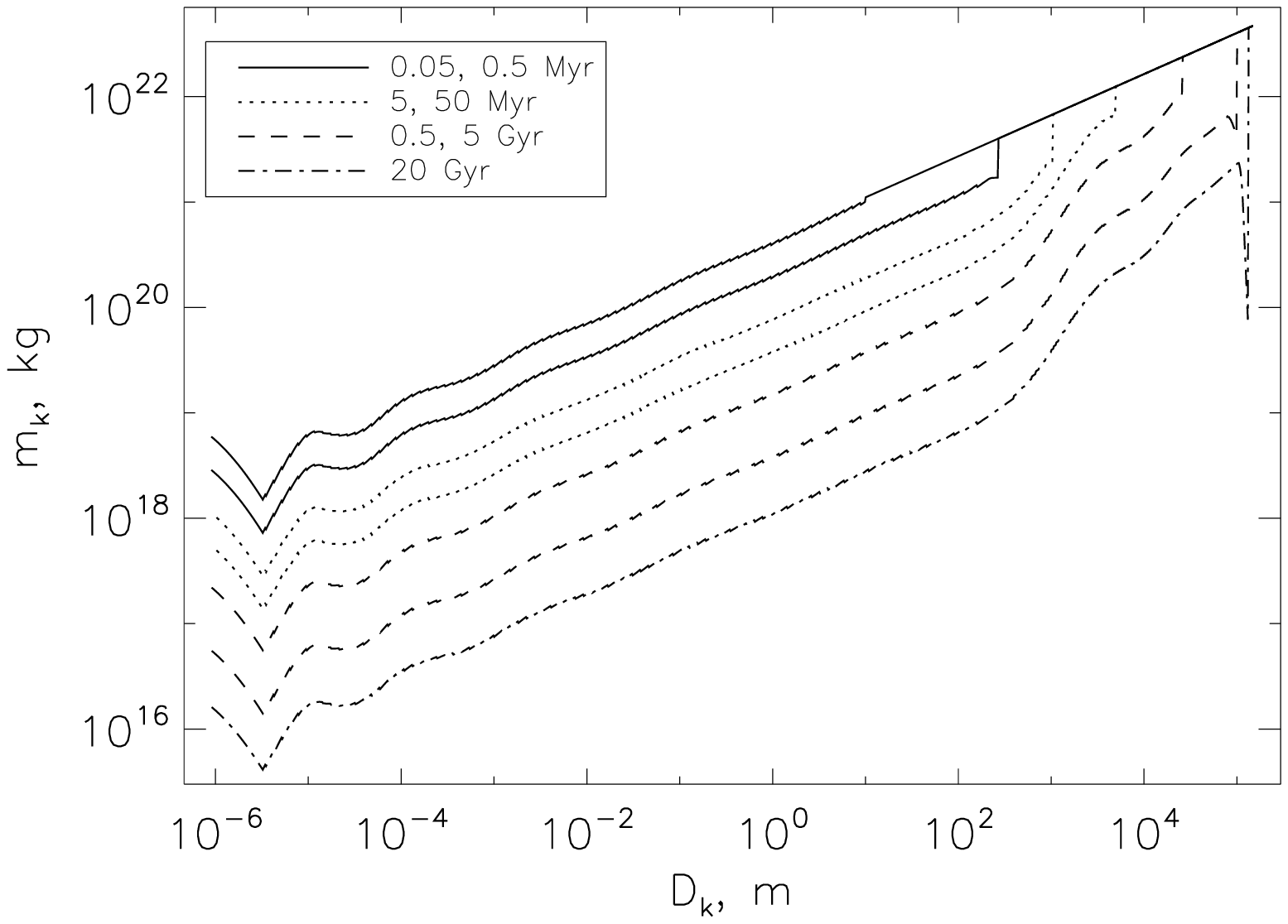,height=1.9in} &
      \hspace{-0.2in} \textbf{(b)} &
      \hspace{-0.5in} \psfig{figure=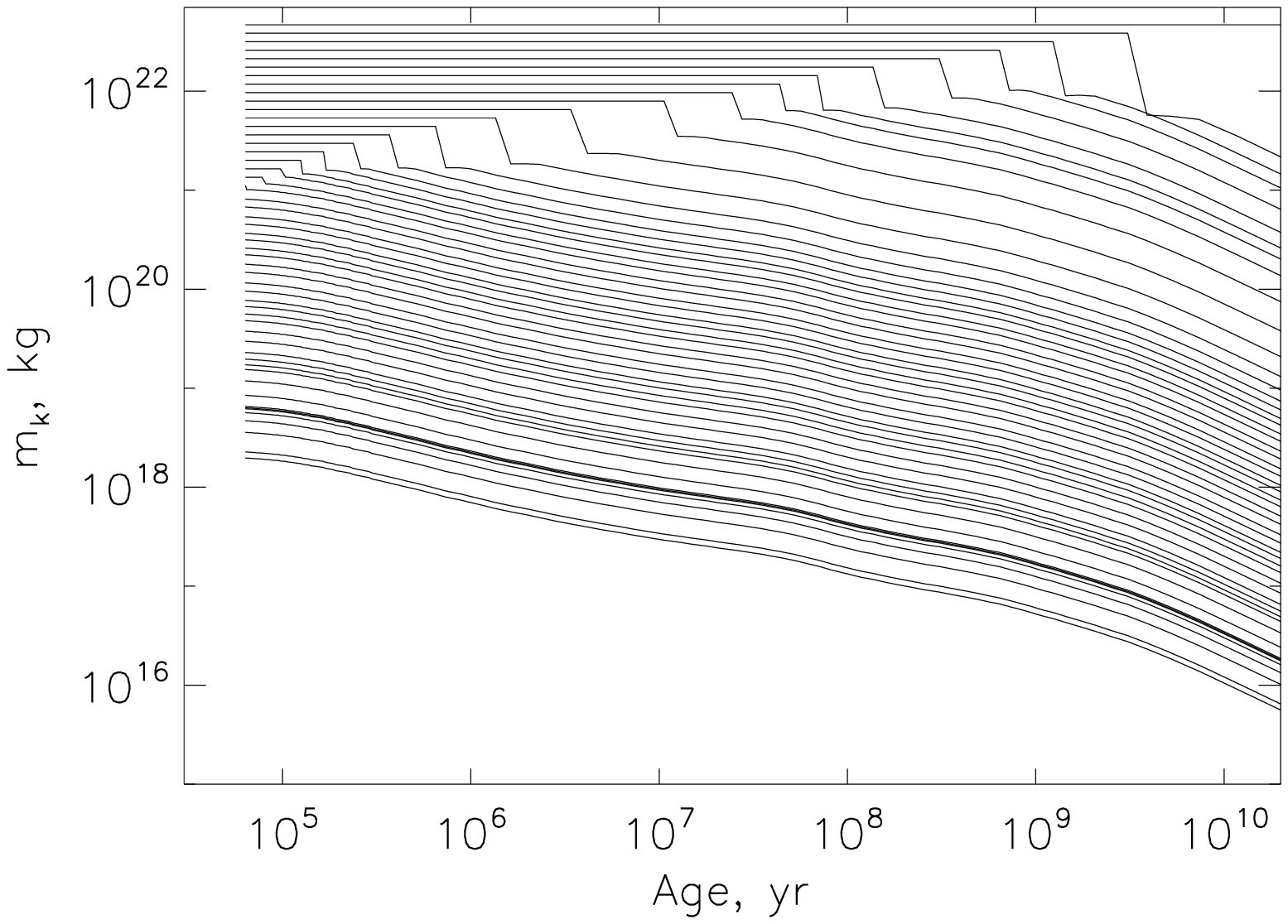,height=1.9in} \\
      \hspace{-0.15in} \textbf{(c)} &
      \hspace{-0.5in} \psfig{figure=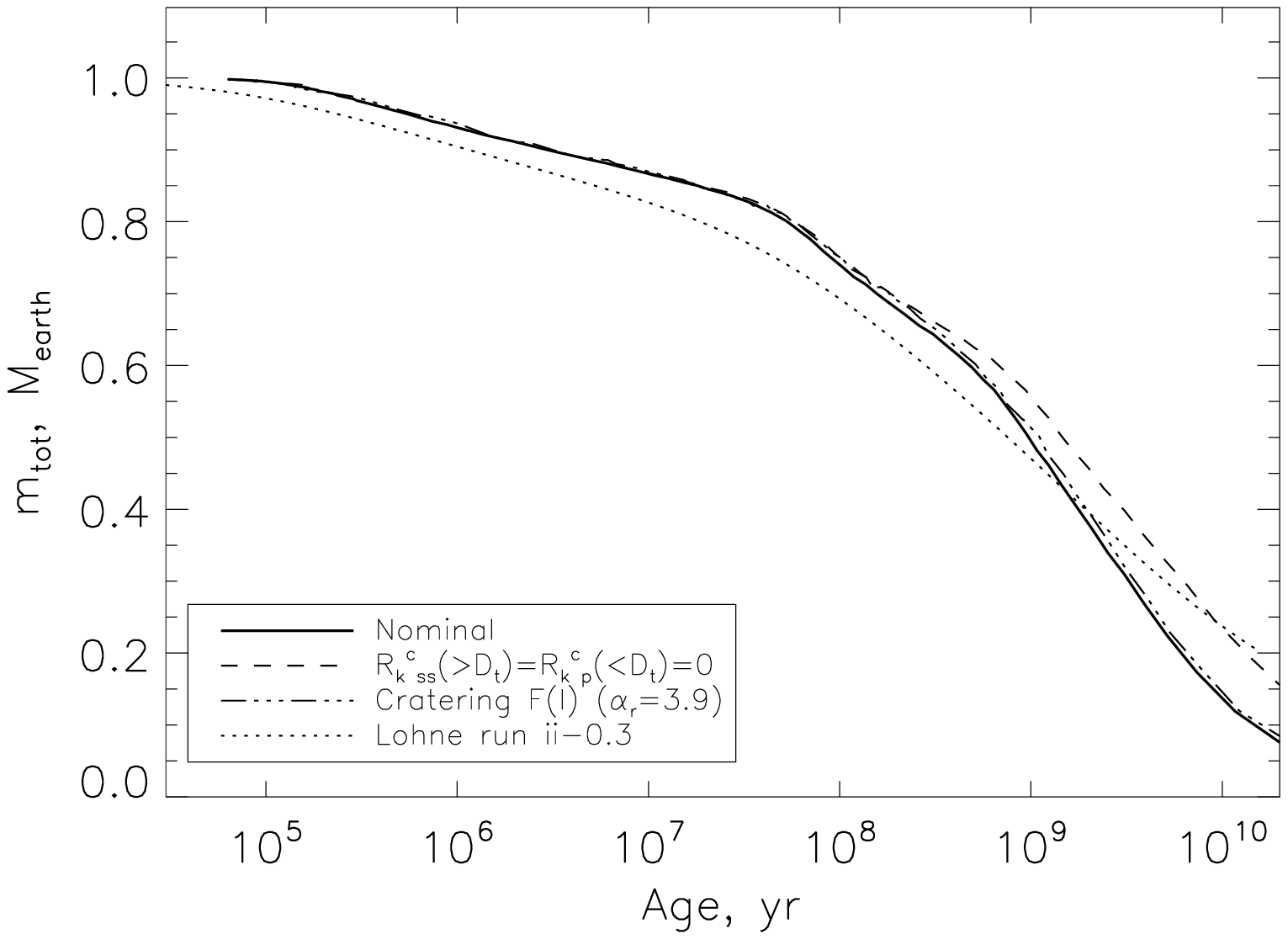,height=1.9in} &
      \hspace{-0.2in} \textbf{(d)} &
      \hspace{-0.5in} \psfig{figure=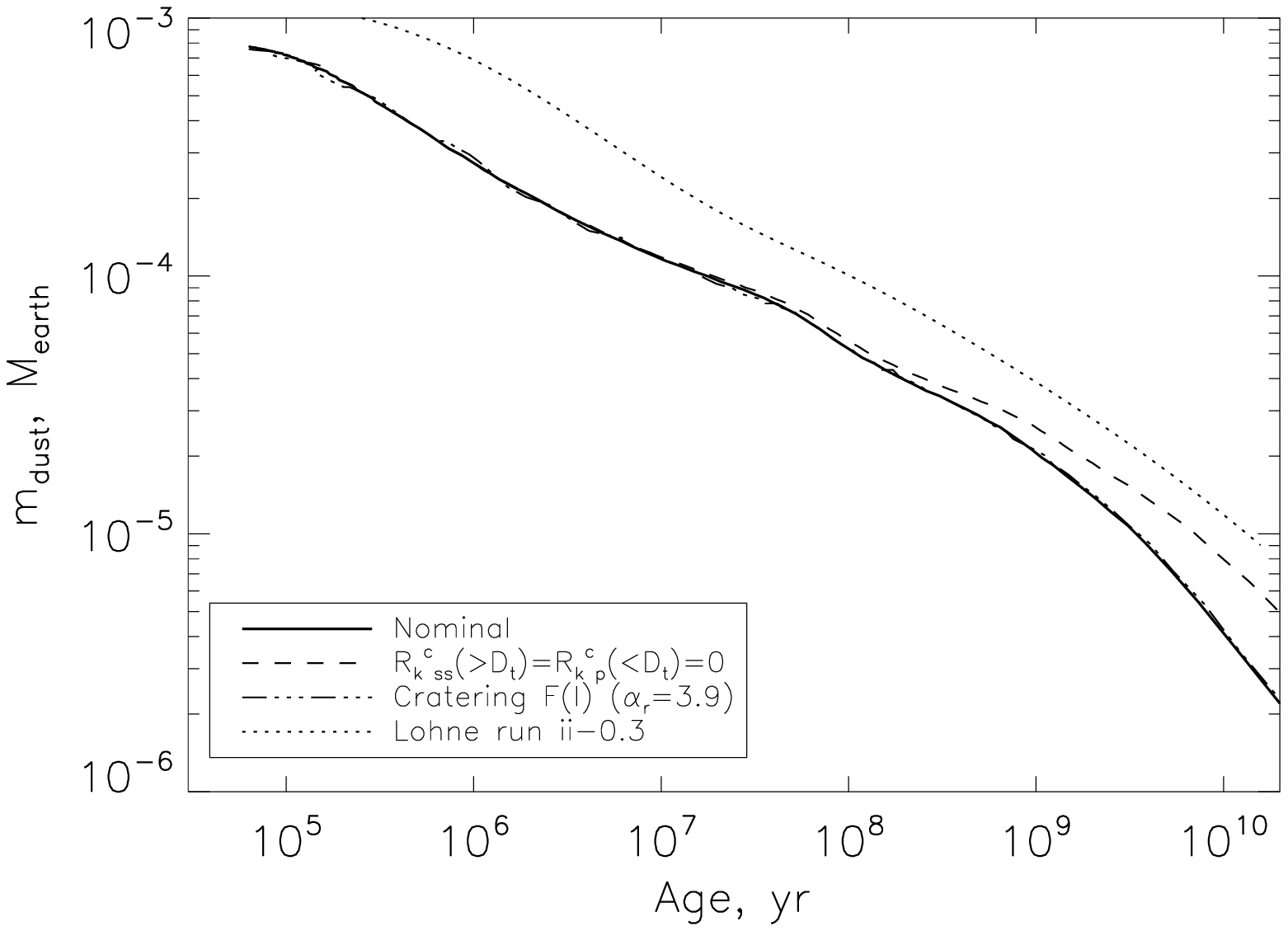,height=1.9in} \\
      \hspace{-0.15in} \textbf{(e)} &
      \hspace{-0.5in} \psfig{figure=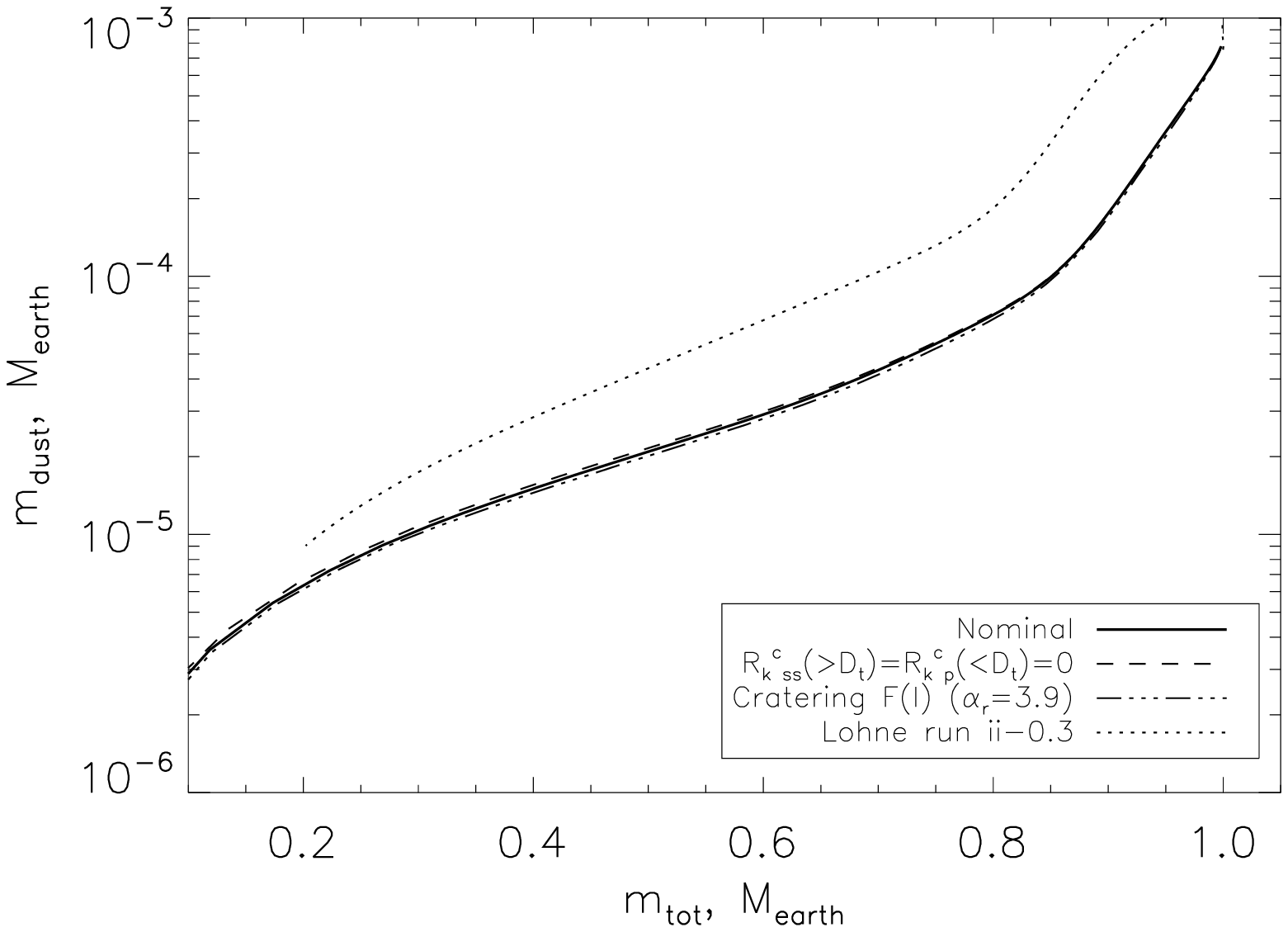,height=1.9in} &
      \hspace{-0.2in} \textbf{(f)} &
      \hspace{-0.5in} \psfig{figure=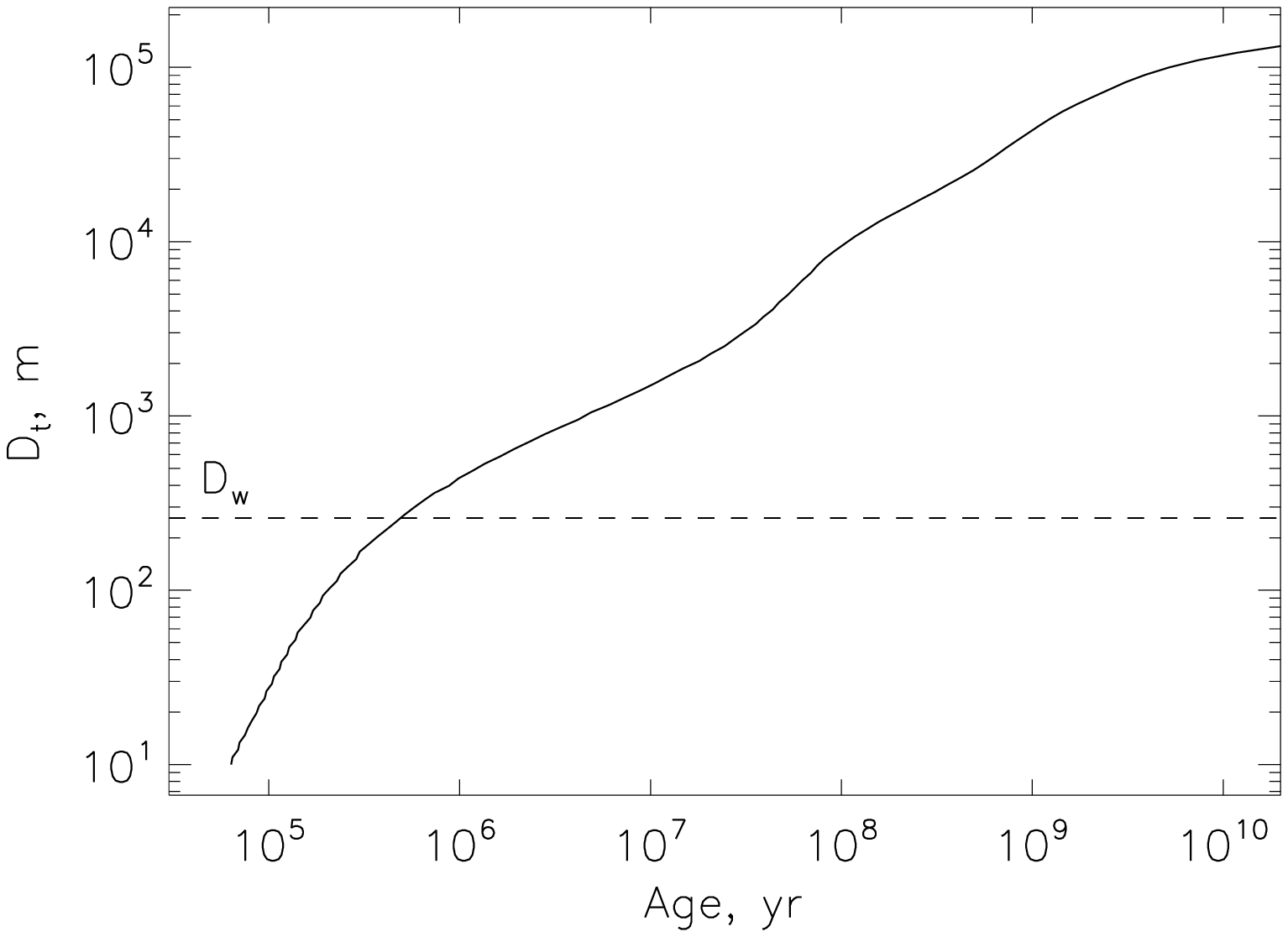,height=1.9in} \\
    \end{tabular}
    \caption{Evolution of a planetesimal belt with the parameters quoted for ACE
    run ii-0.3 of L\"{o}hne et al. (2008).
    For comparison, the results for such an ACE run are reproduced on
    \textbf{(c)}, \textbf{(d)} and \textbf{(f)} with a dotted line (A. Krivov, priv. comm.),
    whereas for \textbf{(a)} and \textbf{(b)} qualitative comparison with figures
    in L\"{o}hne et al. (2008) is possible (see text for details).
    \textbf{(a)} Mass ($m_k$) in different size bins ($D_k$) at times of 0.05, 0.5, 5,
    50, 500 Myr and 5, 20Gyr (compare with L\"{o}hne's Fig.~4 bottom).
    \textbf{(b)} Mass ($m_k$) in 50 different (but equally-logarithmically-spaced) size
    bins as a function of time (compare with L\"{o}hne's Fig.~4 top), where bin size could be
    deduced from plot \textbf{(a)}.
    \textbf{(c)} Total mass as a function of age. 
    \textbf{(d)} Dust mass (in particles $D<2$mm) as a function of age. 
    \textbf{(e)} Dust mass as a function of total mass.
    \textbf{(f)} Transition diameter (from primordial to steady state) as a function of
    age.
    In practise, the size distribution is calculated for a given $D_t$, which is 
    then converted into an age by determining the collision lifetime of objects at the
    transition.
    }
    \label{fig:lohneii-0.3}
  \end{center}
\end{figure*}

To consider how accurately our scheme reflects simulations that solve for
the evolution using the full equation~(\ref{eq:mdotk}) (or its equivalent), we compare
our results with those of the ACE run ii-0.3 of L\"{o}hne et al. (2008).
That run considered the evolution of a 7.5-15AU planetesimal belt with a maximum
eccentricity of 0.3, maximum inclination of 0.15 rad, and largest planetesimals 148km
in diameter, for a Sun-like star.
The primordial distribution had $1M_\oplus$ distributed according to
equation (\ref{eq:nd}) with $\alpha_p=3.61$.
The result of our scheme for these parameters, with $\delta=0.02$ (and so 1279 diameter
bins), is shown in Figure~\ref{fig:lohneii-0.3}.
Figure \ref{fig:lohneii-0.3} should only be compared qualitatively with figures in
L\"{o}hne et al. (2008), since that paper mistakenly used a slightly different
$Q_b=0.014$ J kg$^{-1}$, a factor of 2.5 higher than stated (and used in our nominal runs,
\S \ref{ss:strengthgravity}) (T. L\"{o}hne, priv. comm.).
However, a quantitative comparison is given within Figures \ref{fig:lohneii-0.3}c,
\ref{fig:lohneii-0.3}d and \ref{fig:lohneii-0.3}e, that also show ACE results for
run ii-0.3 with $Q_b=5.6 \times 10^{-3}$ J kg$^{-1}$ (A. Krivov, priv. comm.).
This comparison shows that our scheme mostly reproduces the same features, both
qualitatively and quantitatively, but there are also some qualitative and
quantitative differences that merit further discussion.

For example, the size distribution at specific ages (Figure~\ref{fig:lohneii-0.3}a) compares well
with that of L\"{o}hne et al. (2008).
Although the details of the ripples close to the blow-out limit are not identical,
they are qualitatively similar, and differences are expected due to the models'
different assumptions about collisional outcome and in their treatment of
the dynamics of grains close to this limit.
However, the discontinuity at the transition size is not apparent in the L\"{o}hne et al.
simulations.
Figure~\ref{fig:lohneii-0.3}b illustrates why this discontinuity cannot be physical, since
although the mass evolution of different bins shows generally good agreement, all sizes
undergo a sudden loss of mass at the time they are assumed to reach steady state, whereas
mass evolution should be continuous.
For this reason we do not expect our scheme to reproduce details of the size
distribution at the transition between primordial and steady state,
such as the divot found by Fraser (2009).

The most instructive plots for comparative purposes are those for the evolution of
total mass (Figure~\ref{fig:lohneii-0.3}c) and dust mass (Figure~\ref{fig:lohneii-0.3}d).
Again, these show qualitatively good agreement.
For example, total mass exhibits a slow fall-off with time from 0.1Myr, but speeds
up around 100Myr tending to a fall-off with age $\propto t^{-0.94}$, close to the
inverse fall-off with time predicted above.
However, the total mass lost at 20Gyr is $\sim 92$\% of the initial mass, whereas
the ACE run loses closer to 81\% on the same timescale, indicating that mass loss is 
quantitatively faster in our scheme.
The dust mass shows similarly good qualitative agreement, with dust mass falling off between
1-200Myr $\propto t^{-0.34}$ for our scheme and $\propto t^{-0.41}$ for ACE.
Although it is noticeable that our scheme underpredicts the dust mass by a factor $\sim 2$
at most ages, this is a relatively small difference given the many orders of magnitude in mass
over which the size distribution is being considered. 
However, our scheme also predicts a turnover beyond $\sim 2$ Gyr, tending to dust mass
falling off $\propto t^{-0.97}$.
This is in qualitative agreement with L\"{o}hne et al. who predict an inverse fall-off
with age at late times, once the largest planetesimals have come to
collisional equilibrium, except that the turnover in their plot is not noticeable since it
occurs much later.

Figure~\ref{fig:lohneii-0.3}e shows that the dependence of dust mass on total mass
in our scheme is very similar to that of L\"{o}hne et al. (2008), apart from the
small underprediction of dust mass noted above.
This suggests that a reinterpretation of age could lead to greater agreement between the
schemes.
Remember that the size distribution is worked out for a given $D_t$, and the
corresponding age is then derived by working out the collision rate of objects at the transition
and equating that with the inverse of the age (see Figure~\ref{fig:lohneii-0.3}f).
In Figure~\ref{fig:lohneii-0.3} we consider some of the model parameters that may affect
evolutionary timescales.
First, since collisions with the primordial population may shorten the collision lifetime
at late times (when such a population is not physically realistic), the dashed line
shows what happens when collision rates are calculated assuming that steady state and
primordial particles can only collide with other members of their respective sub-population.
This is implemented by changing the indices over which the collision rate is summed in
equation~(\ref{eq:rkc1}).
As expected this results in slightly longer evolutionary timescales for both total and dust mass.
Notably the evolution is now closer to that found by the ACE simulations in that
85\% of the total mass has been lost at 20Gyr, and the turnover in the dust mass evolution at
2Gyr is no longer present.
We also considered the effect of the redistribution function, by trying a cratering-like
function weighted toward objects of size $D_k$ (with $\eta_{rmax}=1$ and $\alpha_r=3.9$,
see dash-triple-dot line).
This has very little effect on the evolution of either total mass or dust mass, and it can
be concluded that the size distribution in the collisional regime has very little dependence
on the redistribution function (but can be used as a probe of the latter in the P-R drag
regime, \S \ref{s:pr}).

Thus, despite this limited comparison, it seems that our prescription
matches the results of more detailed simulations without the need for
such time-consuming simulations to determine the evolution on long
($>$Gyr) timescales.
Although the 3-phase analytical prescription given in L\"{o}hne et al. (2008)
does likewise, our model has the benefit that we reproduce
more detailed features of the size distribution in the steady state regime
(such as ripples), and moreover it is possible to consider additional
effects such as loss processes like P-R drag (see \S \ref{s:pr}).
We expect that our scheme would replicate the results of other
sets of full simulations (e.g., with different assumptions about collision
outcome), though a detailed comparison would be needed to quantify (and
improve through refining the implementation of the scheme) the level of
agreement.

\section{Steady state with P-R drag}
\label{s:pr}
Despite the uncertainties in the details of how the size distribution transitions from
primordial to steady state, equation~(\ref{eq:epsk}) demonstrates that the small size
end of the distribution is firmly in steady state and so can be treated as such.
Consider now a belt in which both collisions and loss processes are acting.
In this section we will consider the effect of Poynting-Robertson (P-R) drag.
Poynting-Robertson drag is the dominant loss process for dust in
the asteroid belt.

\subsection{Analytical fourth phase}
\label{ss:pr4}
To work out the loss rate from the belt we consider the timescale
for dust particles to migrate from $r_{mid}$ to $r_{mid}-dr/2$.
Clearly, such particles have not yet been removed from the system,
but are then fed into an inner region which should then be considered
in some other way;
our prescription means that we are only considering the size distribution
of particles in the belt region.
The loss rate by P-R drag from the belt for individual particles is
\begin{eqnarray}
  R_{k}^{pr} & = & A_{pr}D_{k}^{-1}, \label{eq:rpr} \\
  A_{pr}     & = & 2.9\times 10^{-6} L_\star/(M_\star \rho dr r_{mid} (1-0.25 dr/r_{mid})),
\end{eqnarray}
where $r_{mid}$ and $dr$ are in AU, $L_\star$ and $M_\star$ in solar units, and $\rho$ in
kg m$^{-3}$, to get a collision rate in yr$^{-1}$ with $D_{k}$ in m.
Thus the loss rate from the $k$-th bin is
\begin{equation}
  \dot{m}_{k}^{-pr} = m_{k} A_{pr} D_{k}^{-1}.
  \label{eq:mdotkpr}
\end{equation}

Because $m_{k} \propto D_{k}^{4-\alpha}$ (from equations \ref{eq:mk} and \ref{eq:nd}),
then in the steady state collisional regime where $m_{k}R_{k}^{c}$ is independent of
$D_{k}$, the loss rate from collisions is 
\begin{equation}
  R_{k}^{c} \propto D_{k}^{\alpha-4}. \label{eq:rckcoll}
\end{equation}
Thus in both the strength and gravity regimes this drops slower than
P-R drag, which falls off as $R_{k}^{pr} \propto D_{k}^{-1}$.
Thus it is possible that at some size, which we call $D_{pr}$ (with a
corresponding index $i_{pr}$), these two are equal, and so
\begin{equation}
  R_{k}^{c}(D_{pr}) = R_{k}^{pr}(D_{pr}).
  \label{eq:dprdef}
\end{equation}
For sizes smaller than $D_{pr}$ the loss is dominated by P-R drag,
whereas for sizes larger than $D_{pr}$ losses are dominated by collisions,
an assumption that will be justified later on.
For some disks $D_{pr}$ is smaller than the size for which particles are
removed by radiation pressure ($D_{bl}$),
and the mass of particles removed by P-R drag is insignificant.
In fact this is necessarily the case in most debris disks that are
bright enough to be detectable (Wyatt 2005).
However, it is not the case for the Solar System's zodiacal cloud, and
it will not be the case for the expected population of fainter extrasolar
debris disks we have yet to discover.

First let us work out what the steady state distribution is for
the size range in which P-R drag is the dominant removal mechanism.
If we assume that the rate of collisional loss in this regime
$\dot{m}_{k}^{-c} \ll \dot{m}_{k}^{-pr}$ and so
can be ignored, then in steady state the equivalent to equation (\ref{eq:ss1})
is
\begin{equation}
  \dot{m}_{ks}^{-pr} = \sum_{i=1}^{i_{pr}} \dot{m}_{is}^{-c} F(k-i),
  \label{eq:mdotkprs}
\end{equation}
where the sum only extends up to $i=i_{pr}$,
since mass loss from $i>i_{pr}$ is assumed not to be collisional.
Since $\dot{m}_{is}^{-c}=C_k$ is approximately independent of size
in the collisional regime (i.e., $C_k=C$), this can be taken outside the sum
giving the steady state distribution as
\begin{equation}
  m_{ks} = A_{pr}^{-1}CD_{k} \sum_{i=1}^{i_{pr}} F(k-i).
  \label{eq:mkspr}
\end{equation}

Thus the size distribution in the P-R drag regime has a slope
that is set by the redistribution function modified by an additional factor
of $D_{k}$.
This prescription for the distribution is continuous at $D_{pr}$ because
at that location the sum in equation~(\ref{eq:mkspr}) is unity and so the
mass calculated in the P-R drag regime at that location is
$m_{ks}(D_{pr}^{-}) = C/R_{k}^{pr}$, which is the same as that calculated
in the collisional regime $m_{ks}(D_{pr}^{+}) = C/R_{k}^{c}$.
For the redistribution function defined in \S \ref{ss:redistrib} by $\alpha_{r}$,
the sum works out to be
$\sum_{i=1}^{i_{pr}} F(k-i) \propto (D_{pr}/D_{k})^{\alpha_r-4}$, which means
that 
\begin{equation}
  m_{ks} \approx A_{pr}^{-1}CD_{pr} (D_{k}/D_{pr})^{5-\alpha_r}.
  \label{eq:mksprf}
\end{equation}
Or in terms of equation~(\ref{eq:nd}) with subscript $pr$,
\begin{equation}
  \alpha_{pr} = \alpha_r-1.
  \label{eq:alphapr}
\end{equation}
This is important because it means that the size distribution of small particles
in the P-R drag regime is an indirect measure of the slope in the redistribution
function.

Since $\alpha_r < 4$ the size distribution in the P-R drag regime is depleted
in small particles.
The main consequence of this is that the rate of collisional losses decreases
toward smaller particles.
This is because the mass loss rate due to collisions given by equation~(\ref{eq:mdotk_dk})
applies regardless of whether the slope in the size distribution $\alpha$ is set
to make the exponent in this equation equal to zero (as is the case in the collisional
steady state regime).
Given that $m_k \propto D_k^{4-\alpha}$ this means that, for particles that are 
small enough to be in the strength regime,
\begin{equation}
  R_k^c \propto D_k^{3-\alpha-(1-\alpha)a/3}.
  \label{eq:rkcnew}
\end{equation}
So in the P-R drag regime where $\alpha$ is set by the redistribution function,
for the parameters assumed here, $R_k^c \propto D_k^{3.8-0.9\alpha_r}$.
Thus, as we have assumed that $\alpha_r <4$, collisional loss rates increase with
size in the P-R drag regime, meaning that
collisional losses can be ignored for all particles smaller than $D_{pr}$, justifying
the earlier assumption.

\subsection{Numerical size distribution}
\label{ss:prnum}
The steady state distribution, including the influence of P-R drag, can be found
numerically by extension of method 2 (\S \ref{sss:method2}).
Again we equate mass loss with mass gain in each bin, to derive a revised
version of equation~(\ref{eq:mks_num2})
\begin{equation}
  m_{ks} = C_k/(R_k^c+A_{pr}D_k^{-1}),
  \label{eq:mks_num3}
\end{equation}
where $C_k$ is again given by equation~(\ref{eq:ck}).

\begin{figure}
  \begin{center}
    \vspace{-0.0in}
    \begin{tabular}{cc}
      \hspace{-0.2in} \textbf{(a)} &
      \hspace{-0.5in} \psfig{figure=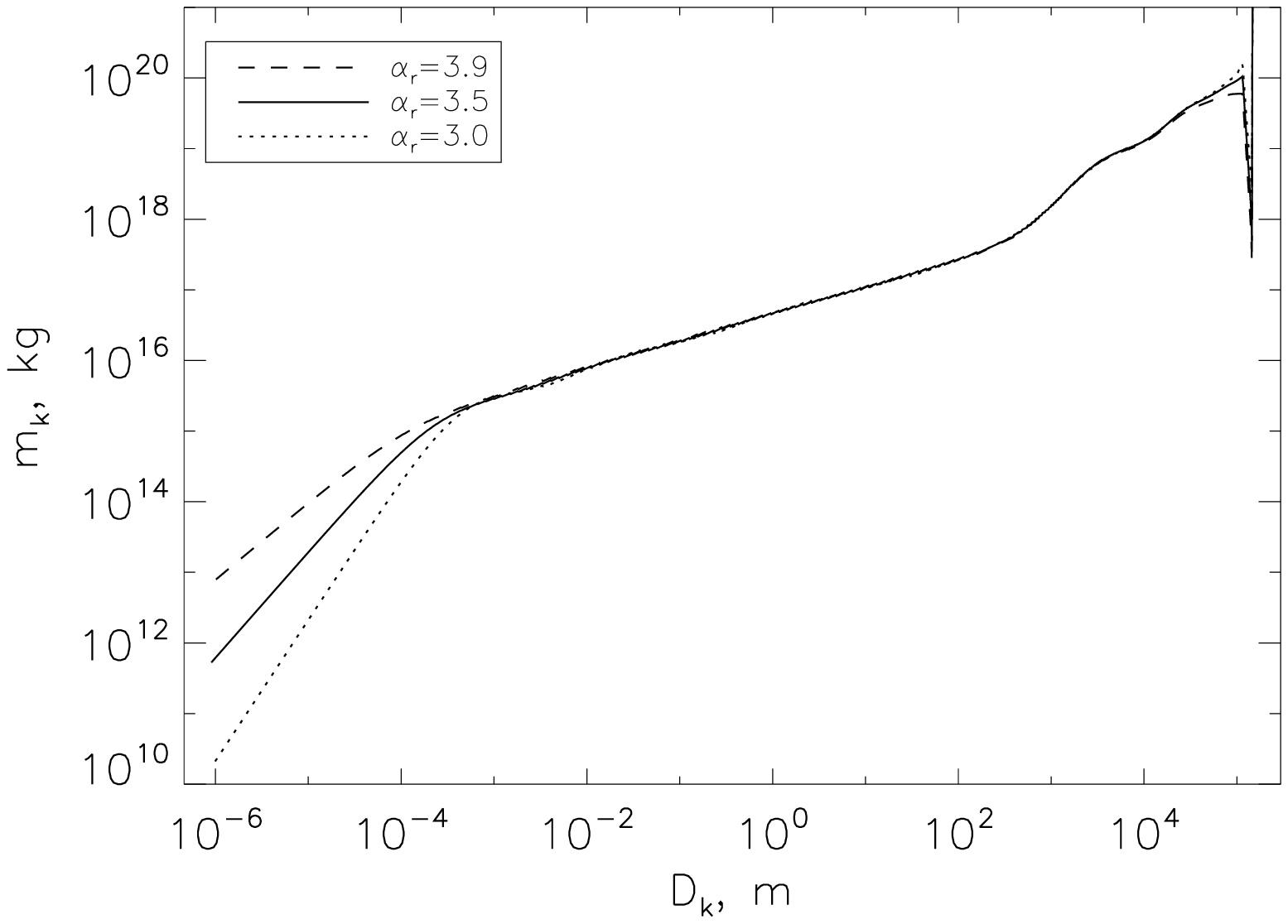,height=2.1in} \\
      \hspace{-0.2in} \textbf{(b)} &
      \hspace{-0.5in} \psfig{figure=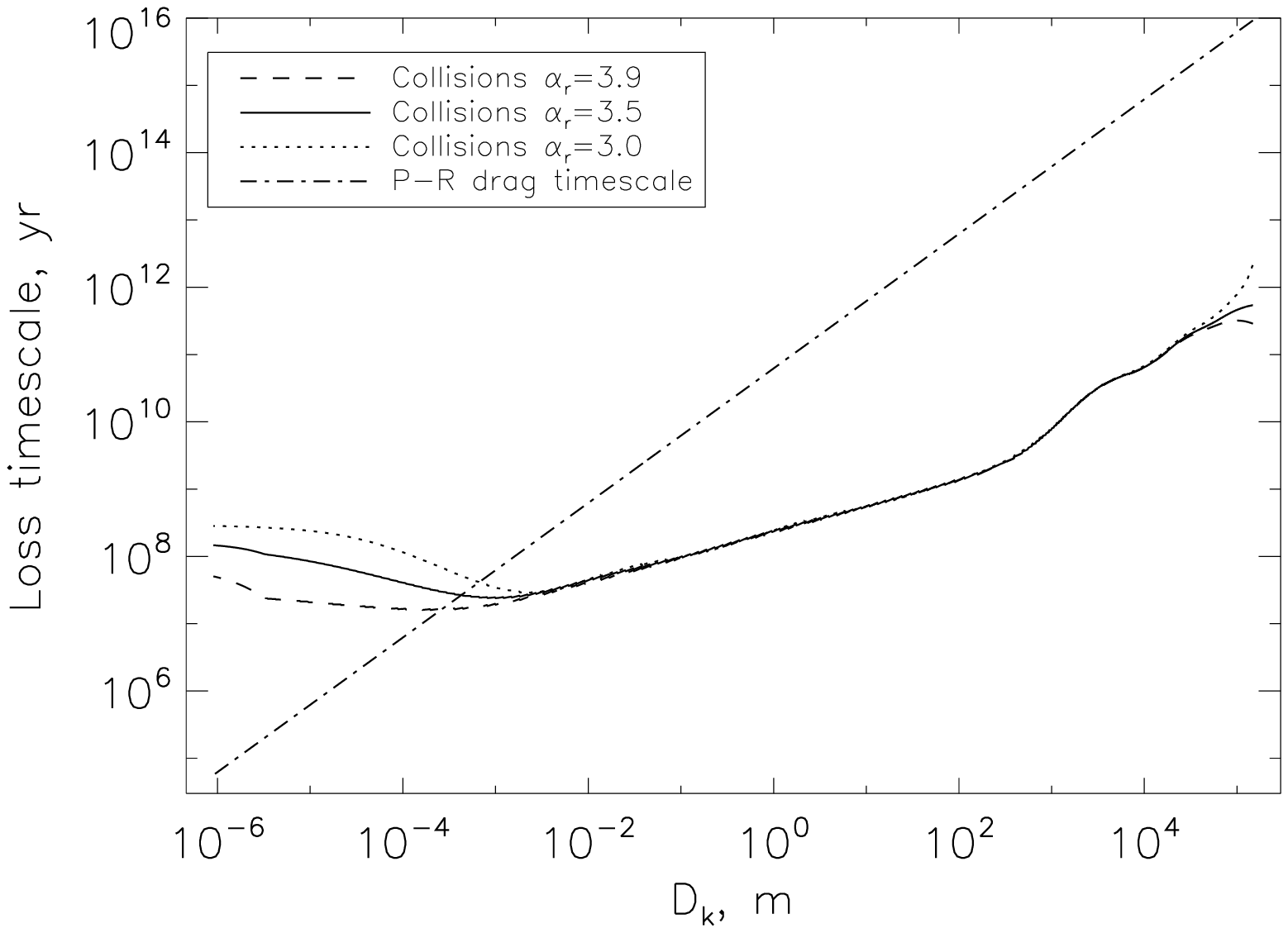,height=2.1in} \\
      \hspace{-0.2in} \textbf{(c)} &
      \hspace{-0.5in} \psfig{figure=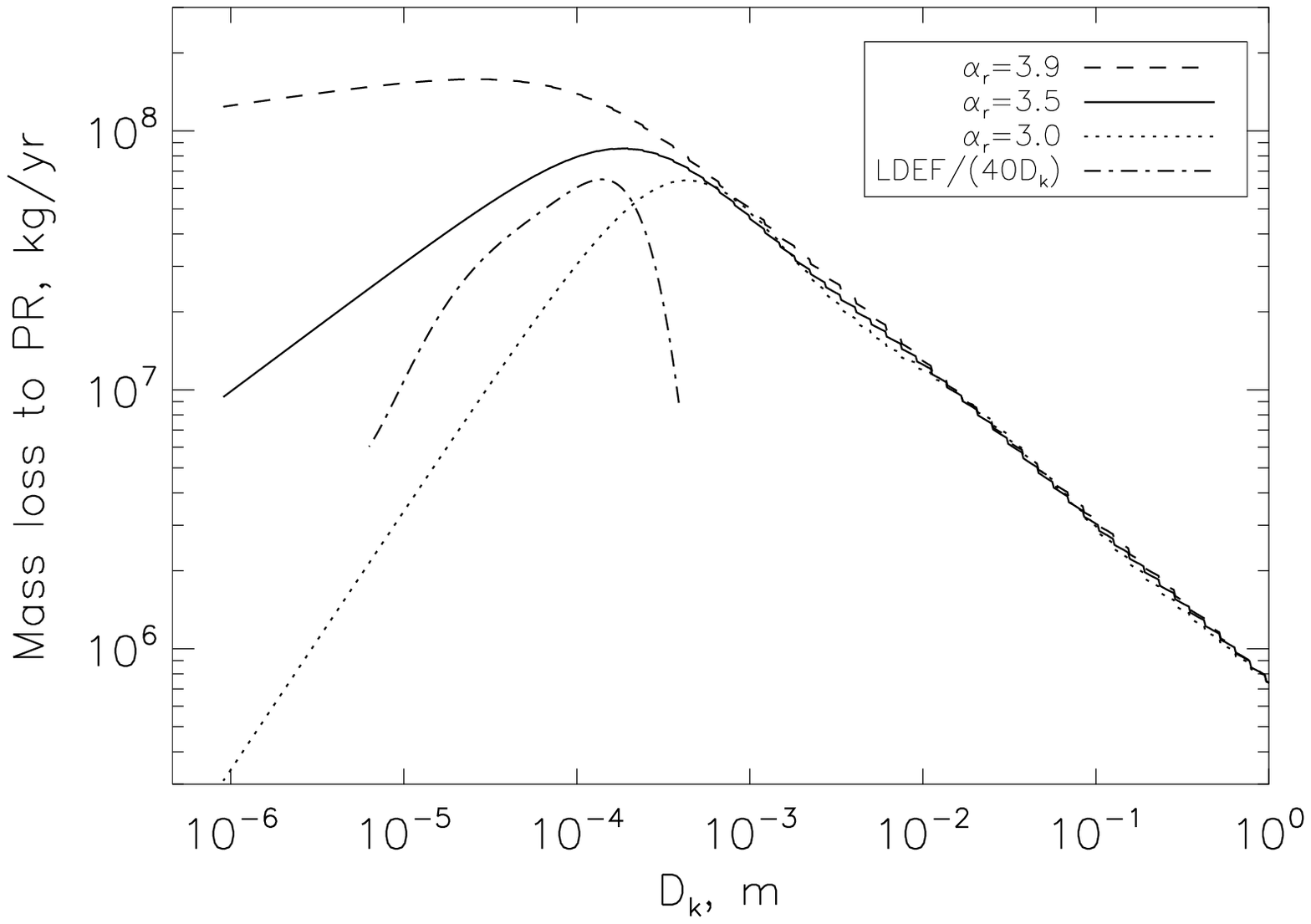,height=2.1in} \\
    \end{tabular}
    \caption{Steady state size distribution of a planetesimal belt with the same
    parameters as assumed in run ii-0.3 of L\"{o}hne et al. (2008), but including loss due
    to P-R drag.
    Different assumptions about the slope in the redistribution function,
    $\alpha_r=3.0,3.5,3.9$ are shown on all plots with dotted, solid and dashed lines,
    respectively.
    The evolutionary age was set to give the same mass of 1m particles for the
    different assumptions, corresponding to an evolutionary age of 290, 540 and 2100 Gyr
    for the different $\alpha_r$ respectively.
    \textbf{(a)} Mass ($m_k$) in different size bins ($D_k$).
    \textbf{(b)} Loss timescales for particles of different sizes in these
    distributions due to collisions and P-R drag.
    \textbf{(c)} Mass loss rate from the inner edge of the belt due to P-R drag.
    The dash-dot line shows the polynomial fit to the accretion rate of dust onto
    the Earth from the LDEF satellite (Love \& Brownlee 1993)
    scaled up by a factor $1/(40D_k)$ to account approximately for the accretion
    efficiency to recover the size distribution in the vicinity of the Earth.}
    \label{fig:pr}
  \end{center}
\end{figure}

This method is used in Figure~\ref{fig:pr} to illustrate the points
raised in \S \ref{ss:pr4} regarding the resulting distribution.
To simplify the discussion, these simulations use exactly the same
parameters as those used in the comparison with run ii-0.3 of
L\"{o}hne et al. (2008) that was discussed in \S \ref{ss:evoln} and
Figure~\ref{fig:lohneii-0.3}.
The main difference is that here we include the effect of P-R drag,
and we also vary the parameter $\alpha_r$.
Since the effect of P-R drag only becomes noticeable with these parameters
for ages $\gg 10$Gyr, its exclusion from the previous section
(and from L\"{o}hne et al.'s simulations) does not compromise those results.
However, this means that we are considering here the distribution at
$\sim 500$ Gyr ages that are greater than the age of the Universe.
Although it seems ridiculous to consider such ages, this is not a major concern,
since the same features would appear on shorter timescales in belts with different
parameters (e.g., with smaller $D_1$ or $r_{mid}$), or in those started with
low total mass, and effects such as dynamical
depletion can be modelled using collisional evolution timescales longer than the
system age (e.g., Bottke et al. 2005; \S \ref{sss:dyndepl}).

Figure~\ref{fig:pr}a shows how P-R drag causes a turnover in the size distribution
toward smaller particles, and how the slope in the P-R drag regime is
set by the slope of the redistribution function, which is exactly as expected
(remembering that the slope on this figure is expected to be $5-\alpha_r$).
It also shows how insensitive the rest of the size distribution is to the
redistribution function.
It is evident that the ripples in the size distribution at small sizes
have disappeared, which is expected above the blow-out limit, as the loss of
particles just above this limit is no longer controlled by collisions.
However, the turnover at $D_{pr}$ has the potential to cause a ripple in the
size distribution just above this size, which is not seen presumably because the
turnover is a gradual rather than sharp feature.

As for the location of the turnover, Figure~\ref{fig:pr}b shows that the diameter
for which $R_{k}^{c}=R_{k}^{pr}$ is $D_{pr}=660$, 430, 260 $\mu$m for $\alpha_r=3.0$, 3.5, 3.9,
respectively, noting that $R_{k}^{c}$ is calculated from the size distribution
of Figure~\ref{fig:pr}a rather than that with no consideration of P-R drag
(which from extrapolation of the power law in the collisional regime would have
resulted in a smaller value of $D_{pr} \approx 150$ $\mu$m that is independent of
$\alpha_r$).
For all values of $\alpha_r$, this is a factor of 1.9-2.0 times larger than the
sizes obtained from the intersections of power law size distributions describing the 
P-R drag and collisional regimes, and a factor of 1.5-10 times larger than the
peaks in the cross-sectional area distributions.

The final Figure~\ref{fig:pr}c shows the mass loss from the belt due to P-R drag.
Since this mass can be considered as a gain term for a population of particles that
are evolving due to P-R drag we denote this as $\dot{m}_{kpr}^{+}$ which is
$\propto m_{k} D_{k}^{-1}$, and so from equations~(\ref{eq:mk}) and (\ref{eq:nd}) is
also $\propto D_{k}^{3-\alpha}$.
Thus in the P-R drag regime, $D<D_{pr}$, this distribution is set by the redistribution
function
\begin{equation}
  \dot{m}_{kpr}^{+}(D<D_{pr}) \propto D_k^{4-\alpha_r},
\end{equation}
whereas in the collisional regime, $D>D_{pr}$, it is set by the strength law
\begin{equation}
  \dot{m}_{kpr}^{+}(D>D_{pr}) \propto D_k^{(-3-2a)/(6-a)}.
\end{equation}
As mentioned above, this mass is not lost from the system, but continues to evolve
toward the star due to P-R drag, also suffering collisions on the way.
It is beyond the scope of this paper to include the extra radial dimension
needed to consider the evolution of this material properly.
However, we can consider what happens if collisions no longer amend the size
distribution, and the particles also pass a planet onto which they may be accreted.
In this case the size distribution of accreted particles would simply be given by
Figure~\ref{fig:pr}c modified by a size dependent accretion efficiency $P_{acc}$.
Accretion efficiencies may plausibly vary $P_{acc} \propto D_k$ due to the slower
motion of larger particles past the planet (e.g., equation B2 of Wyatt et al. 1999),
though the detailed dynamics may be more complicated (Kortenkamp \& Dermott 1998).

The assumption that collisions can be ignored is likely too crude to be
useful for specific applications;
in particular, collision timescales may be expected to be shorter than P-R drag timescales for
particles larger than $D_{pr}$.
However, it is notable that this size distribution is not so dissimilar to that
derived for the accretion of dust onto the Earth by LDEF (Love \& Brownlee 1993),
a scaled version of which is also shown on Figure~\ref{fig:pr}c.
This suggests that, with further work, it may be possible to learn about the
redistribution function from such observations.
For example, if the collisionless and $P_{acc} \propto D_k$ assumptions held, then
Figure~\ref{fig:pr}c shows that the LDEF accretion rate for $<200$ $\mu$m particles
would imply a redistribution function with $\alpha_r$ slightly below 3.5, perhaps with
a turnover to a smaller value below a few 10s of $\mu$m.
Gr\"{u}n et al. (1985) also noted that the size distribution in the
zodiacal cloud at 1AU may be indicative of the redistribution function, from
a consideration of collisional gains and losses due to collisions and P-R drag,
though these authors favoured the interpretation that the zodiacal cloud is not
in steady state to altering the redistribution function from their
assumed value of $\alpha_r=3.5$.

In fact the redistribution function is poorly constrained.
This distribution for the break-up of large objects can be measured from asteroid
families and is known to depend on impact energy, partly due to geometrical
considerations (Tanga et al. 1999), but is typically found to range from
$\alpha_r=3.5-4.5$.
Collisional experiments of (smaller) rocks (Fujiwara et al. 1989) give distributions
that have $\alpha_r=2.8-5.8$ (though 3.7-4.0 is more common).
However, it is noted that the distributions can turn over to lower values of $\alpha_r<3.5$
at fragment sizes smaller than 1mm (e.g., Flynn et al. 2009),
i.e., in exactly the size range of interest, also changing to even lower values for
$<100$ $\mu$m.
Thus, although the distribution implied by Figure~\ref{fig:pr}c would appear to have a
relatively small value of $\alpha_r$, this is not without precedent within the context
of previous studies, especially for this size range.

\subsection{Evolution in P-R regime}
\label{ss:prnumevol}
The evolution in the P-R drag regime was discussed in Dominik \& Decin (2003) who
found that the dust luminosity should decrease $\propto 1/t^2$ in this regime.
We find that the situation is substantially more complicated than the prescription
given by these authors, since we treat an evolving size distribution in which
the numbers of particles of different sizes evolve differently.
Nevertheless, for certain assumptions we recover the same dependence
for the evolution of luminosity.

To consider the evolution in the P-R drag regime, we first need to determine the
evolution of $D_{pr}$.
Here we will consider the situation in which the largest particles are in equilibrium,
so that the mass in all bins above $D_{pr}$ are being depleted with age as $1/t$.
The implications for Figure~\ref{fig:pr}b are that the collision lifetimes of all
particles (above $D_{pr}$) increase $\propto t$, and so the turnover evolves to larger sizes.
Here we ignore the effect of the turnover on the calculation of the collision time at
$D_{pr}$;
this means that we miscalculate $D_{pr}$, but by a fraction that should be
the same at all ages for a given $\alpha_r$, and so should not affect conclusions about
the evolution.
Thus equating collision and P-R loss rates we find that 
\begin{equation}
  D_{pr} \propto t^{(6-a)/(3+2a)},
  \label{eq:dpr}
\end{equation}
or $D_{pr} \propto t^{1.58}$ for the parameters of our simulation.

For particles larger than $D_{pr}$, their number continues to deplete inversely with
age.
However, for particles that are now in the P-R drag regime, the increase of $D_{pr}$
with age means that they are being replenished by a population for which the size
at which its bottom end is truncated is also increasing with age.
The number of particles in the P-R drag regime thus depletes much faster
\begin{equation}
  m_{ks}(D<D_{pr}) \propto t^{[\alpha_r(6-a)-30]/(3+2a)},
  \label{eq:mksdpr}
\end{equation}
or $\propto t^{-2.79}$ for the parameters of our simulation with $\alpha_r=3.5$.
The evolution that is observed then depends on what particle sizes are being
probed.
For example, the evolution of dust mass defined as the mass in $D<2$mm particles would
be expected to fall off inversely with age at early times when $D_{pr} \ll 2$mm, but tend
to a fall off $\propto t^{-2.79}$ by the time $D_{pr} \gg 2$mm,
and likewise for similar definitions of dust mass.

However, for certain observations the contribution to the emission from the disk is
dominated by the total cross-sectional area in the distribution.
For the distributions considered here, with $\alpha_r<4$, that area is dominated by
particles of size $\sim D_{pr}$.
Although this size evolves in a way that depends on the strength law (equation~\ref{eq:dpr}),
it turns out that the evolution of total cross-sectional area does not, 
since $\sigma_{tot} \propto m_k(D_{ref})D_{pr}^{(3+2a)/(a-6)}$, where $m_k(D_{ref})$
is the mass in particles at some fixed reference diameter $D_{ref}$ that is above
$D_{pr}$.
In the regime we are considering in which the largest particles have reached collisional
equilibrium $m_k(D_{ref}) \propto t^{-1}$, and so from equation~(\ref{eq:dpr}) we find that 
$\sigma_{tot} \propto t^{-2}$ as predicted by Dominik \& Decin (2003).
Neverthless, all observations are prone to bias toward probing certain particle
sizes (see, e.g., Fig.~5 of Wyatt \& Dent 2002), and so one would expect all observations
to tend to fall off at a rate given by equation~(\ref{eq:mksdpr}) on long enough timescales,
and it is important to consider the particle sizes contributing to a particular observation
to understand how disk brightness is expected to evolve.

\begin{figure}
  \begin{center}
    \vspace{-0.0in}
    \begin{tabular}{cc}
      \hspace{-0.2in} \textbf{(a)} &
      \hspace{-0.5in} \psfig{figure=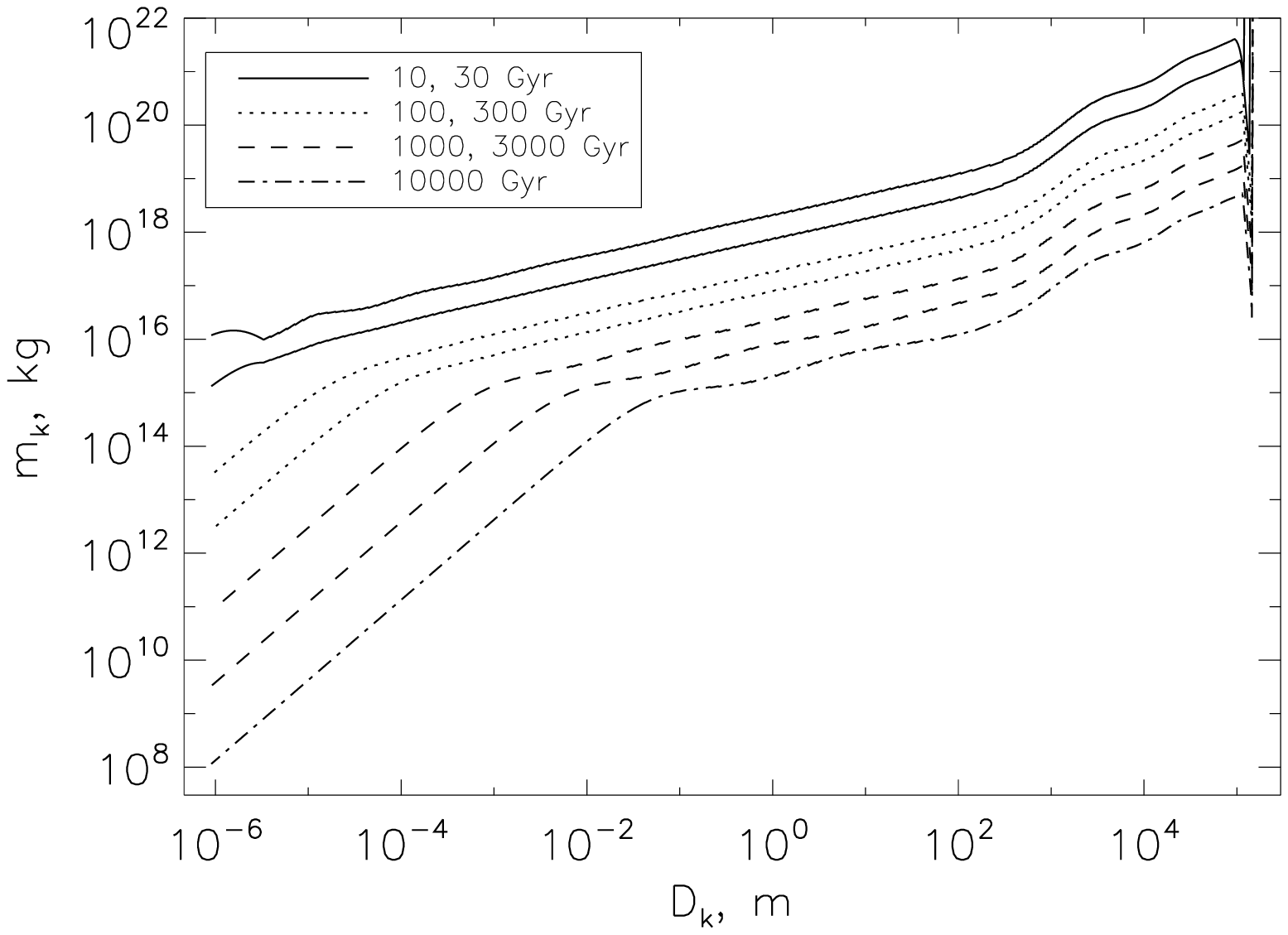,height=2.1in} \\
      \hspace{-0.2in} \textbf{(b)} &
      \hspace{-0.5in} \psfig{figure=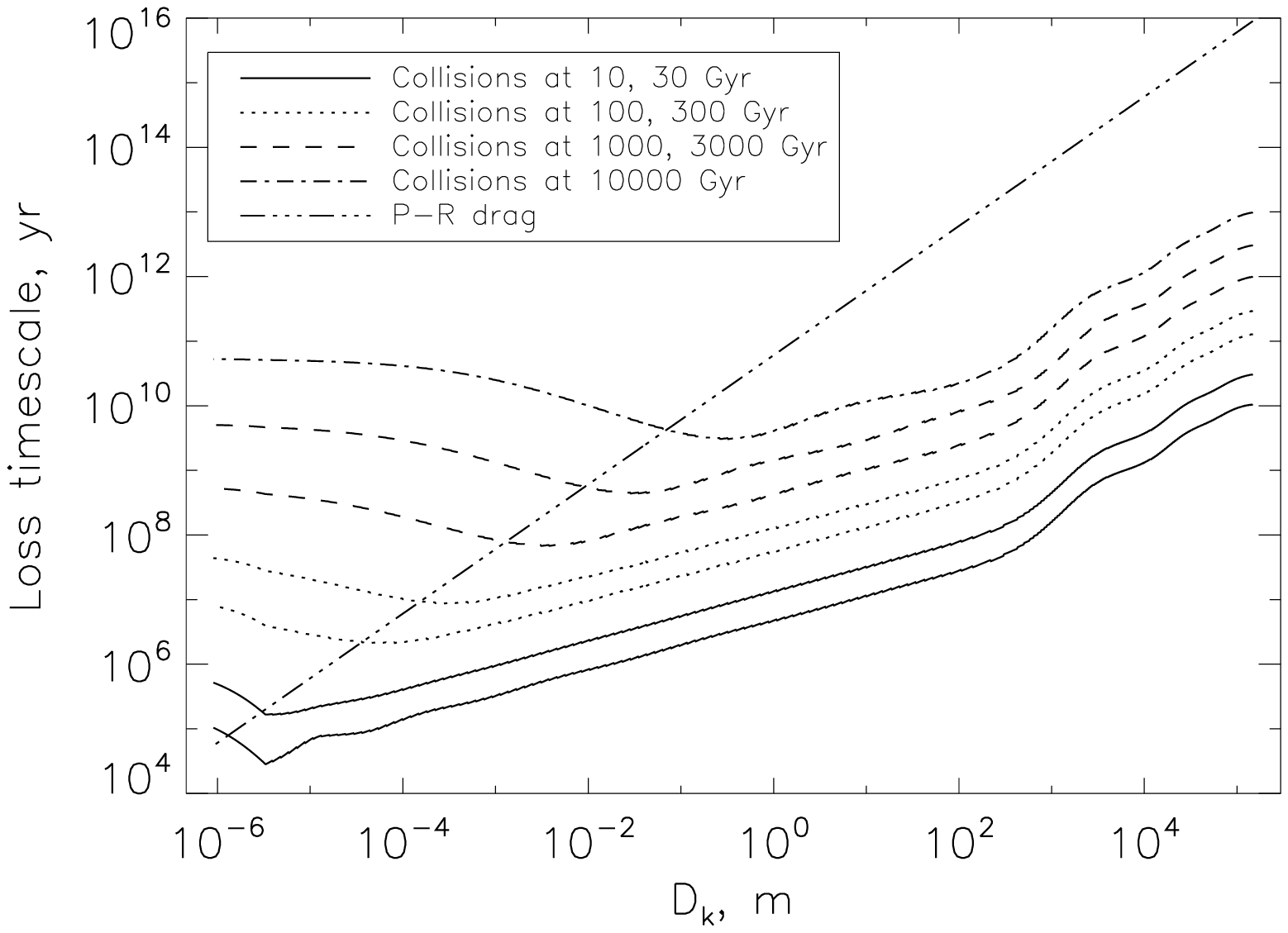,height=2.1in} \\
      \hspace{-0.2in} \textbf{(c)} &
      \hspace{-0.5in} \psfig{figure=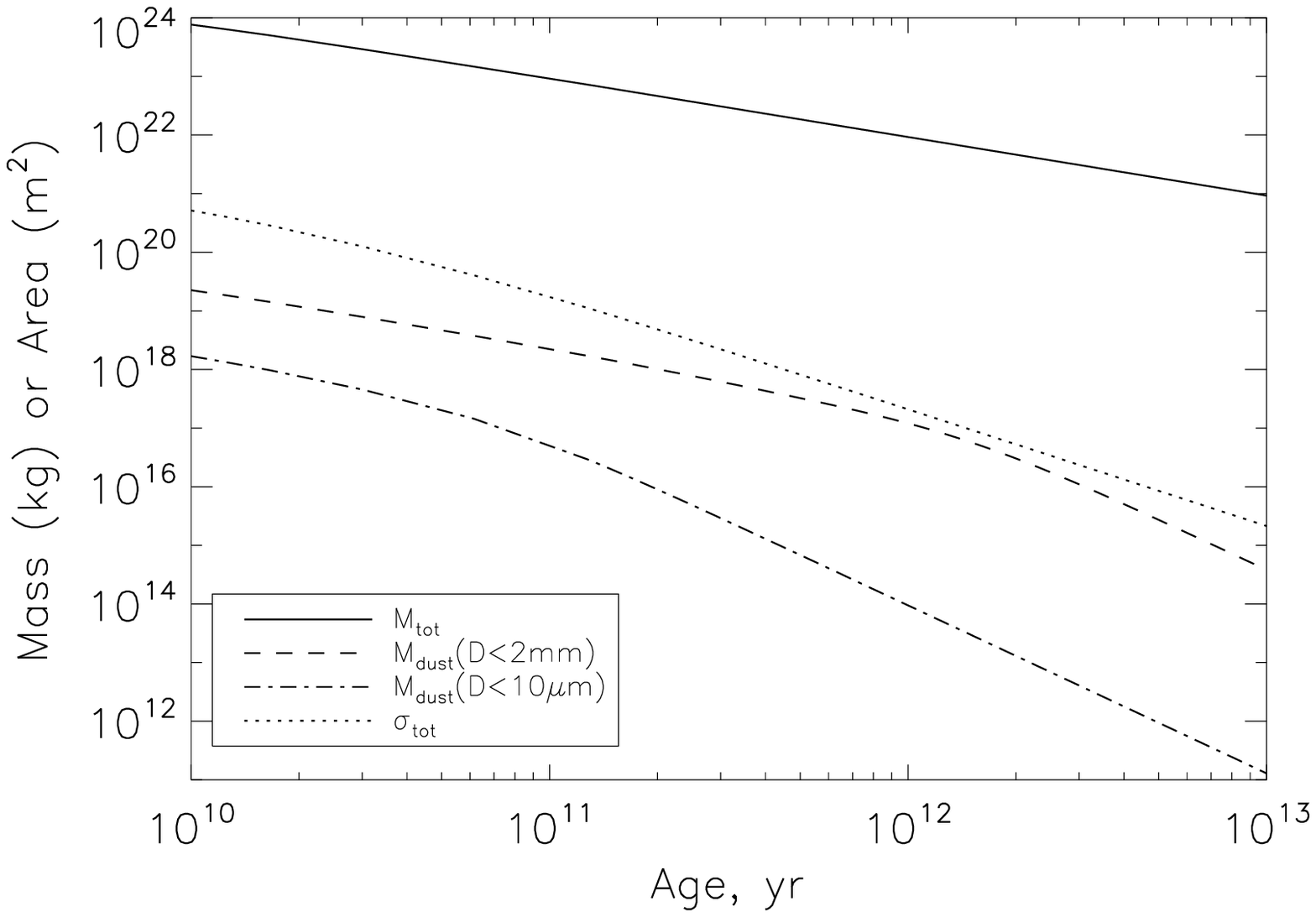,height=2.1in} \\
    \end{tabular}
    \caption{Evolution of the planetesimal belt considered in Figure~\ref{fig:lohneii-0.3}
    on timescales long enough for P-R drag to affect the evolution of the dust.
    Although these timescales are unrealistically long for this parameter set,
    shorter timescales may apply for different belt parameters, and dynamical depletion
    can result in evolutionary timescales that are longer than the system age
    (\S \ref{sss:dyndepl}).
    \textbf{(a)} Mass ($m_k$) in different size bins ($D_k$) at ages of 10, 30, 100, 300,
    1000, 3000, 10000 Gyr.
    \textbf{(b)} Loss timescales for particles of different sizes at these ages
    due to collisions and P-R drag.
    \textbf{(c)} Evolution of total mass, dust mass (both defined as mass in particles
    smaller than 2 mm, and smaller than 10 $\mu$m), and cross-sectional area.}
    \label{fig:prevol}
  \end{center}
\end{figure}

The above points are illustrated in Figure~\ref{fig:prevol}, which shows the evolution
from an age of 10Gyr of the planetesimal belt considered in Figure~\ref{fig:lohneii-0.3}, but
including P-R drag at all ages (note that $\alpha_r=3.5$ in these simulations).
Figure~\ref{fig:prevol}a shows how the size distribution at 10Gyr is unaffected by P-R drag
(justifying its exclusion from the simulations shown in Figure~\ref{fig:lohneii-0.3}),
but that soon after this the turnover due to P-R drag appears at a size that gets larger
with age.
It is also evident that the ripple above $D_{pr}$ that was noted to be absent from
Figure~\ref{fig:pr} is now apparent at late ages.
This is because at larger values of $D_{pr}$, particles at this transition have lower values
of $X_c$, i.e., they are destroyed by particles that are much smaller relative to themselves
than those with smaller values of $D_{pr}$, which means that the turnover appears more like a
truncation to them.
Figure~\ref{fig:prevol}b shows how the turnover size is indeed set at all ages by the location
where collision and P-R drag timescales are equal.
This justifies the analytical work in the last few paragraphs, which also provides excellent
quantitative agreement with the evolution of mass and cross-sectional area seen in
Figure~\ref{fig:prevol}c;
indeed at late times the simulations find $m_{tot} \propto t^{-1.0}$, $m_{dust} \propto t^{-2.8}$,
$\sigma_{tot} \propto t^{-2.0}$.

A word of caution is necessary on the evolution implied by equations~(\ref{eq:dpr}) and
(\ref{eq:mksdpr}), which is that these apply when the largest objects in the distribution are
in collisional equilibrium and so are being depleted $\propto 1/t$.
This may not be the case for belts that were simply born with low mass, or have undergone
dynamical depletion (like the asteroid belt), or have enhanced levels of drag (e.g. \S \ref{ss:sw}).
For such systems it would be necessary to consider the rate of depletion of particles with
sizes above $D_{pr}$, and substitute that into the above analysis in place of the $\propto 1/t$
dependence assumed there.

\subsection{Summary of evolutionary stages}
\label{ss:sumevol}
To summarise, if we ignore the ripples in the distribution, then the evolution of the size
distribution shown by the
runs of Figs.~\ref{fig:lohneii-0.3} and \ref{fig:prevol} show four main stages of evolution
that are summarised in Table~\ref{tab:tab1}.

\begin{table}
  \begin{center}
    \caption{Evolutionary stages for the belt discussed in Figs.~\ref{fig:lohneii-0.3}
    and \ref{fig:prevol}, with the ages over which the stage lasted, the phases present
    in the size distribution (P=Primordial, S=Strength regime, G=Gravity regime, PR=P-R drag
    regime), and the dependences on time of total disc mass, dust mass and total
    cross-sectional area.}
    \begin{tabular}{cccccc}
      \hline
      Stage & Ages          & Phases & $m_{tot}$        & $m_{dust}$          & $\sigma_{tot}$      \\
      \hline
      0     & 0             & P      & -                & -                   & -                   \\
      I     & $<0.5$Myr     & S-P    & flat             & flat                & flat                \\
      II    & 0.5Myr-2Gyr   & S-G-P  & gradual          & $\propto t^{-0.34}$ & $\propto t^{-0.34}$ \\
      III   & $2-30$Gyr     & S-G    & $\propto t^{-1}$ & $\propto t^{-1}$    & $\propto t^{-1}$    \\
      IV    & $>30$Gyr      & PR-S-G & $\propto t^{-1}$ & $\propto t^{-2.8}$  & $\propto t^{-2}$    \\
      \hline
    \end{tabular}
    \label{tab:tab1}
  \end{center}
\end{table}

The evolution of planetesimal belts with different initial parameters would be expected to show similar
behaviour.
However, there would be quantitative differences.
For example, the ages of the transitions between the stages would depend on the initial parameters, since they
correspond (e.g., Table~\ref{tab:tab2}) to the collisional lifetime of objects of
size $D_w$ (transition I-II), and of size $D_1$ (transition II-III),
and the age at which the collisional lifetime of objects of size $D_{bl}$ is equal to their P-R drag lifetime
(transition III-IV).
The fall-off of dust mass and area with age during stage II also depends on the primordial size distribution and
on the dispersal law in the gravity regime (see equation~43 of L\"{o}hne et al. 2008),
and the dust mass in stage IV depends on the dispersal law in the strength regime and the redistribution function
(equation~\ref{eq:mksdpr}).
There may also be qualitative differences.
For example, if the primordial distribution contained no particles in the gravity regime,
i.e., if $D_1<D_w$, then there could be no "G" phase in the size distribution, and stage II would
be skipped.
Similarly, if the primordial distribution was made up of different power laws, then there
would be different phases associated with the timescale for collisional equilibrium to reach the transitions
between the primordial power laws.
And it was already noted that it may be possible for P-R drag to become important during stage II in which
case an evolutionary stage involving the 4 phases "PR-S-G-P" would be reached.

\section{Steady state with generalised loss processes}
\label{s:gen}
The method of \S \ref{s:pr} for considering the effect of P-R drag on the steady
state size distribution of planetesimal belts undergoing a collisional cascade
can be applied to other loss processes in a similar manner.
In particular, if the loss process is described by
\begin{equation}
  \dot{m}_{k}^{-l} = m_{k} R_{k}^{l},
  \label{eq:mkl}
\end{equation}
then its effect on the steady state size distribution can be determined
numerically by extending method 2 (\S \ref{sss:method2}) to solve
\begin{equation}
  m_{ks} = C_k/(R_{k}^{c} + R_{k}^{l}).
\end{equation}
Here we consider several loss processes, and their qualitative effect
on the size distribution (and its evolution), but defer a detailed treatment
to later work.
Treatment of other loss processes such as Yarkovsky forces and gas drag
are also possible within this framework, as are erosive loss processes,
such as sublimation and sputtering, by invoking the gain term
$\dot{m}_{k}^{+g}$ in equation~(\ref{eq:mdotk}) (since loss from one
size bin results in gain into another).
However, such considerations are again beyond the scope of this paper.

\subsection{Power law loss rate}
\label{ss:powerlaw}
Consider the action of a loss process that can be defined by
equation~(\ref{eq:mkl}) with
\begin{equation}
  R_{k}^{l} \propto D_k^{-\gamma_l},
  \label{eq:rkl}
\end{equation}
which is a generalisation of loss by P-R drag for which $\gamma_l=1$.

\subsubsection{Large $\gamma_l$}
\label{sss:largegaml}
As long as $\gamma_l$ is large enough, the analytical treatment is identical
to that for P-R drag, in that there is a transition at the size $D_l$ for which
\begin{equation}
  R_{k}^{c}(D_l)=R_{k}^{l}(D_l).
  \label{eq:dldef}
\end{equation} 
For $D<D_l$ the removal of particles is dominated by the loss process, while
for $D>D_l$ removal is dominated by collisions.
A consideration of the analogous Figure~\ref{fig:pr}b shows that
\textit{large enough} here means $\gamma_l > \gamma_{lim}$ where
$\gamma_{lim}$ is determined from a comparison of the slope in the loss regime
(equation~\ref{eq:rkl}) with that in the collisional regime (equation~\ref{eq:rckcoll}).
Assuming the transition occurs in the strength regime (i.e., $D_l < D_w$),
equation~(\ref{eq:alphaa}) means that
\begin{equation}
  \gamma_{lim} = 4 - \alpha_a = (3-3a)/(6-a),
\end{equation}
which for $a=0.3$ means $\gamma_l>0.37$.
Thus the evolution is qualitatively similar, and the size distribution in the
loss regime is still indicative of the redistribution function, except that
the index is now 
\begin{equation}
  \alpha_l = \alpha_r - \gamma_l.
  \label{eq:alphal}
\end{equation}
The mass loss from the belt by loss processes is
$\dot{m}_{ks}^{-l} \propto D_k^{\gamma_{lim} - \gamma_l}$ for $D>D_l$,
but is still $\dot{m}_{ks}^{-l} \propto D_k^{4-\alpha_r}$ for $D<D_l$.
Assuming the largest particles are in collisional equilibrium so that particles
in the collisional regime are being depleted $\propto t^{-1}$, then by equating
the rate of collisional loss $R_{k}^{c} \propto D_{k}^{\alpha_a-4}t^{-1}$ with
the rate from loss processes (equation~\ref{eq:rkl}), we find that the transition
size follows
\begin{equation}
  D_l \propto t^{\frac{1}{\gamma_l-\gamma_{lim}}}.
  \label{eq:dl}
\end{equation}
For similar reasoning to that used to derive equation~(\ref{eq:mksdpr}), this
means that the mass in small particles below the transition evolves
\begin{equation}
  m_{ks}(D<D_l) \propto t^{\frac{\alpha_r-4-2(\gamma_l-\gamma_{lim})}{\gamma_l-\gamma_{lim}}}.
  \label{eq:mksdl}
\end{equation}

\subsubsection{Dynamical depletion}
\label{sss:dyndepl}
However, if $\gamma_l<\gamma_{lim}$, then it is not small particles
that are most affected by loss processes, but particles above $D_l$.
This is the situation for most dynamical loss processes for which
loss rates are independent of size and so $\gamma_l=0$.
The effect on the size distribution is more complicated in this case.

For example, consider the belt of Figure~\ref{fig:lohneii-0.3} that has
evolved to $t_{dep}=0.5$Gyr without dynamical mass loss, but then undergoes
impulsive and significant dynamical depletion by a factor $f_{dep}$.
This would cause the size distribution to be retained, but 
to drop at all sizes by the same fraction.
This would mean that additional evolutionary time would be needed for the smallest objects
in the primordial distribution ($\sim 27$km in diameter) to come to
collisional equilibrium and continue to evolve.
In the meantime the size distribution in the portion that was originally
in steady state ($D<27$km) would be retained since the depletion would
not have affected the balance of mass gain or loss from such bins.

This example illustrates two ways in which dynamical depletion can affect 
the size distribution of what appears to be primordial at late times:
some parts of the distribution may represent a scaled down version of the
original primordial distribution, whereas others may represent the steady state
distribution attained during an earlier high mass phase of evolution, even though
their collisional lifetime is longer than the current system age.
This point was recognised by Bottke et al. (2005), who showed how
dynamical depletion can be modelled by considering collisional evolution
that takes place over timescales longer than the system age.
This is in agreement with our analysis, since we expect the size distribution
of the above example following the depletion to look like that expected for
a primordial distribution that is scaled down by $f_{dep}$ then evolved
for a time $1/f_{dep}$ times longer than the current system age.
As noted by Bottke et al., it is not possible to infer $f_{dep}$
from the current distribution (of the asteroid belt in their application),
i.e., we cannot determine if the original distribution was very massive and
short-lived, or less massive and longer-lived.
However, it is possible to model the evolution of the size distribution
for depletion by a known amount $f_{dep}$ at a known time $t_{dep}$.

\subsection{Realistic radiation forces}
\label{ss:realbeta}
The prescription for the effect radiation forces in the preceding sections is an
approximation of the true physics.
For example, the truncation of the size distribution by radiation pressure, which
occurs for particles with $D<D_{bl}$ is implemented by simply not including such
bins in the analysis.
However, a more realistic approach would be to include radiation pressure as a
loss process with an associated (short, i.e., less than orbital) timescale for
bins below $D_{bl}$.
Thus it would be possible to account for the possibility that the blow-out population
collides with bound grains (Krivov et al. 2000), and for a gradual rather than
sharp truncation due to a finite eccentricity of the parent particles (which means that
$\beta>0.5$ particles can remain bound) and/or due to a range of particle properties.
It was already noted that the increase in eccentricity for particles close to $D_{bl}$
is not accounted for in this model, but again this could be incorporated using an
appropriately modified intrinsic collision probability $P_{ik}$ in
equation~(\ref{eq:rkc1}) (e.g., Krivov et al. 2005; Wyatt et al. 2010).

Another approximation is the inverse linear dependence of the P-R drag loss rate
on particle size. 
In fact $R_{k}^{pr} \propto \beta$, and although $\beta \propto 1/D$ for
large black body particles, this function peaks at 0.1-1 $\mu$m particles
(i.e., at particle sizes that are comparable to the dominant wavelength of
stellar radiation), turning over for smaller particles and tending to $\beta$
independent of size for particles $\ll 0.1$ $\mu$m (e.g., Gustafson 1994).
For particles around high luminosity stars this turnover is inconsequential, since
it occurs at sizes below $D_{bl}$.
However, for low luminosity stars it is so important that there may be no particles
that can be removed by radiation pressure, i.e., $\beta<0.5$ for all particle sizes
(e.g., Sheret et al. 2004).
The consequence of this is that, in the absence of other loss processes, P-R drag is
the only removal mechanism for dust from the cascade, and so it is important to
consider the effect of the turnover in $\beta$ vs $D$ on the steady state size
distribution.
A reconsideration of equation~(\ref{eq:mksprf}) shows that the size distribution
in this regime should really be
\begin{equation}
  m_{ks} \propto D_k^{4-\alpha_r}/\beta.
  \label{eq:mkspr4}
\end{equation}
This goes some way to explaining the shape of the size distribution seen
in Fig.~4 of Reidemeister et al. (2010), since equation~(\ref{eq:mkspr4}) would predict
a distribution of drag dominated particles that is fairly flat in the 0.04-10 $\mu$m size
range with a dip at $\sim 0.4$ $\mu$m.

\subsection{Stellar wind}
\label{ss:sw}
Although the discussion has focussed on radiation forces, it is notable
that P-R drag only becomes important for debris disks that are so low in density
that they are no longer detectable with current instrumentation
(Dominik \& Decin 2003; Wyatt 2005).
However, stellar wind forces act in exactly the same way as radiation
forces.
Their pressure component (which is analogous to radiation pressure) can usually
be ignored, but their drag component (which is analogous to P-R drag) can significantly
exceed P-R forces (e.g., Plavchan et al. 2005; Reidemeister et al. 2010).
Thus stellar wind drag can be considered as an enhanced form of P-R drag using the same
analysis but with a constant $A_{sw}$ in equation~(\ref{eq:rpr}) that means its
effect on the size distribution can become apparent at much younger evolutionary
ages (i.e., for much higher, and potentially detectable, disc masses) than its
radiation counterpart.
As per the discussion of \S \ref{ss:realbeta}, it may also be the dominant
loss mechanism for the belts of low luminosity stars.
No change in the qualitative picture of disc evolution presented in
\S \ref{ss:sumevol} is necessary, except to note that the turnover in
the size distribution may become apparent in earlier evolutionary phases
(e.g., in phase II as already noted in \S \ref{ss:sumevol}).

\section{Conclusion}
\label{s:conc}
In this paper we presented a new scheme for determining the shape of the size distribution,
and its evolution, for collisional cascades of planetesimals undergoing destructive collisions
and loss processes like Poynting-Robertson drag.
This scheme, outlined in \S \ref{s:col}, considers mass gain and loss in different size bins, and
treats the steady state portion of the cascade by solving for the assumption that $\dot{m}_{k}=0$,
and assumes that larger objects retain their primordial distribution.
Many results can be obtained analytically, providing insight into the origin of the shape of
the size distribution.
For example, for particles in the steady state collisional regime this prescription was shown
to lead to mass loss rate being independent of size in logarithmic size bins;
thus the steady state size distribution is independent of mass gain into the bins in this
regime.
However, perhaps most notably this scheme provides an efficient and readily implemented
numerical method for calculating the size distribution.
Although the scheme as presented assumes a simple scale independent redistribution function
that defines the average mass distribution of fragments produced in collisional destruction of
objects of a given size, and considers only the size distribution in one radial bin, it may be
expanded to more complex redistribution functions, and to include a radial dimension (or to
include dimensions of orbital elements).

Our scheme, and its application to consider the evolution of the size distribution
in the collision-dominated regime, was shown to reproduce qualitatively all of the features
found by more detailed (and computationally more time-consuming) models that step size
distributions forward in time.
Such features in the size distribution include ripples above the radiation pressure blow-out
limit, the change in slope at the transition from strength to gravity regime, the ripple above
this transition, and the transition to primordial regime.
For evolution, the scheme reproduces the initially slow fall-off of dust mass with
time $\propto t^{-0.34}$ followed by steeper decline $\propto t^{-1}$, as well as an
evolution of total disk mass that is much flatter in the initial phases but also tends to
$\propto t^{-1}$ at late ages.

One of the main purposes of the paper was to consider the effect of loss processes on the
steady state size distribution, which was done in detail for P-R drag in \S \ref{s:pr}.
This was found, both analytically and confirmed numerically, to cause a turnover at small
sizes to a size distribution that is set by the redistribution function.
This means that information about the redistribution function may be recovered by measuring
the size distribution of particles undergoing loss by P-R drag.
This was illustrated by comparison with the size distribution of particles accreted
by the Earth, though it was noted that further work on the collisional evolution
of material inside the source belt and on the accretion efficiency as a function of particle
size would be needed before definitive conclusions on the redistribution function could be
reached.
The evolution of the size distribution when P-R drag is included is summarised in \S \ref{ss:sumevol}
and Table~\ref{tab:tab1}.
Notably it is found that although cross-sectional area drops with age $\propto t^{-2}$ in the
PR-dominated regime, dust mass falls $\propto t^{-2.8}$, underlining the importance of
understanding which particle sizes contribute to an observation when considering how disk
detectability evolves.

Other loss processes are readily incorporated into this scheme.
In \S \ref{s:gen}, we also discussed loss rates that can be defined by
a power law in particle size, including dynamical depletion, a more
realistic prescription for radiation loss rates that are not simply 
inversely proportional to paticle size, and stellar wind drag.
Steep power law loss rates lead to size distributions and evolutions that are
directly analogous to P-R drag (but with modified exponents), but loss rates that
are less biased to small particles are harder to treat.
However, the application to dynamical depletion agrees with the results in the literature.
The way in which P-R drag becomes independent of size for the smallest particles
causes a ripple and flattening of this distribution at such sizes in the P-R drag
regime.
The application to stellar wind drag results in distributions and evolutions that
are the same as those for P-R drag, but if this effect becomes important at all, it
does so at higher disc masses meaning that it has the potential to be relevant
for disks that are detectable with current technology.

There are now several surveys for debris disks that show how
disk brightness drops with age (see review in Wyatt 2008), and further
surveys are currently underway (e.g., Matthews et al. 2010).
Interpretation of these surveys requires efficient schemes so that
populations of large numbers of debris disks can be modelled
(e.g., Wyatt et al. 2007b; L\"{o}hne et al. 2008).
As such surveys push to lower disk masses, it is also important
to be able to incorporate consideration of loss processes.
The results presented here suggest that when disks drop to a level at which
P-R drag becomes important, which is close to the limit of detectability
with Spitzer (see Fig. 4 of Wyatt et al. 2007a), their brightness begins to
drop rapidly, as fast as $\propto t^{-2.8}$.
Such a fast fall off is also expected for disks affected by stellar wind drag,
which may occur for disks well above the current detectability threshold.
In short, this paper provides an efficient model with which to interpret
observations of the debris disks of nearby stars, and their evolution,
including in regimes where loss processes are important for small (or large)
particles.

\section*{Acknowledgments}
The authors are grateful to the Isaac Newton Institute for Mathematical Sciences
in Cambridge, where the initial work on this paper was carried out during the Dynamics
of Discs and Planets research programme, and to Torsten L\"{o}hne for
discussions on his ACE simulations.
We are also grateful to the reviewers for their careful reading of the manuscript
and to Alexander Krivov for providing the results of the ACE simulation presented
in Figure 2.
Mark Booth is grateful to UK STFC and to CSA for financial support.


\end{document}